
%


\hoffset=2cm
\overfullrule=0pt
\font\tenbf=cmbx10
\font\tenrm=cmr10
\font\tenit=cmti10
\font\ninebf=cmbx9
\font\ninerm=cmr9
\font\nineit=cmti9

\font\eightrm=cmr8
\font\eightit=cmti8

\hsize=5.0truein
\vsize=7.8truein
\parindent=15pt


\newcount\sectionno	\sectionno=0
\newcount\appno		\appno=0
\newcount\subsectionno	\subsectionno=0
\newcount\subsubsectionno	\subsubsectionno=0
\newcount\eqnum		\eqnum=0
\newcount\refno   	\refno=0
\newcount\footno   	\footno=0

\def\eq{\global\advance\eqnum by1 \eqno{(\mainid\the\eqnum})}
\def\eqlabel#1{\eq {\xdef#1{\mainid\the\eqnum}}}

\newwrite\refs
\def\jr#1{{\nineit#1}}
\def\vol#1{{\ninebf#1}}
\def\aref#1{\global\advance\refno by1
 \immediate\write\refs{\noexpand\item{\the\refno.}#1\hfil\par}}
\def\ref#1{\aref{#1}\the\refno}
\def\refname#1{\xdef#1{\the\refno}}

\def\section#1{\vglue 7pt
 \global\advance\sectionno by1 \subsectionno=0 \eqnum=0
 \message{<\the\sectionno>}
	\leftline{\tenbf \the\sectionno. #1}
	\nobreak\vglue 5pt\nobreak\def\mainid{\the\sectionno.}}

\def\subsection#1{\vglue 7pt
 \global\advance\subsectionno by1 \subsubsectionno=0
 \message{<\mainid\the\subsectionno>}
	\leftline{\tenit \mainid\the\subsectionno\ #1}
	\nobreak\vglue 5pt\nobreak}

\def\subsubsection#1{\vglue 7pt
 \global\advance\subsubsectionno by1
 \message{<\mainid\the\subsectionno.\the\subsubsectionno>}
	\leftline{\tenrm\mainid\the\subsectionno.\the\subsubsectionno\ #1}
	\nobreak\vglue 5pt\nobreak}

\def\appendix#1\par{\vskip0pt plus.1\vsize\penalty-250
 \vskip0pt plus-.1\vsize\bigskip\vskip\parskip
 \global\advance\appno by1 \subsectionno=0 \eqnum=0
 \def\mainid{\ifcase\appno\or A.\or B.\or C.\or D.\or E.\or F.\or G.\fi}
	\message{<\mainid>}
	\leftline{\tenbf #1}
	\nobreak\vglue 5pt}

\def\foot#1{\global\advance\footno by1\def\footid{
\ifcase\footno\or a\or b\or c\or d\or e\or f\or
g\fi}\nobreak\footnote{$^\footid$}{\eightrm #1}}

\def\KM{Ka\v c-Moody\ }
\def\La{\Lambda}

\def\Tr{{\rm Tr}}
\def\Zb{{\bf Z}}
\def\alb{\bar{\alpha}}
\def\alv{\alpha^{\vee}}
\def\alvb{{\bar\alpha}^{\vee}}
\def\al{\alpha}
\def\and{\quad {\rm and}\quad}
\def\at{\tilde{a}}
\def\bt{\tilde{b}}
\def\bullet{.}
\def\com{\quad,\quad}
\def\ct{\tilde{c}}
\def\ep{\epsilon}
\def\frac#1#2{{\textstyle{#1\over #2}}}
\def\gb{\bar{g}}
\def\gh{\hat{g}}
\def\hf{\frac12}
\def\hh{\hat{h}}
\def\hb{\bar{h}}

\def\lab{\bar{\lambda}}
\def\la{\lambda}
\def\l{\left}
\def\mh{\hat{m}}

\def\mod#1{~~({\rm mod}~#1)}
\def\mub{\bar{\mu}}
\def\nh{\hat{n}}

\def\r{\right}
\def\su#1{\hat{su}(#1)}
\def\so#1{\hat{so}(#1)}

\def\thb{\bar{\theta}}

\def\wt{\tilde{w}}


\nopagenumbers
\baselineskip=10pt
\rightline{\tenrm LAVAL-PHY-25/91\ ,\ LETH-PHY-9/91}
\rightline{\tenrm (September 1991)}
\vglue 5pc
\baselineskip=13pt
\headline{\ifnum\pageno=1\hfil\else%
{\ifodd\pageno\rightheadline \else \leftheadline\fi}\fi}
\def\rightheadline{\hfil\eightit
Field identification in nonunitary diagonal cosets \quad\eightrm\folio}
\def\leftheadline{\eightrm\folio\quad \eightit Mathieu et al\hfil}
\voffset=2\baselineskip
\centerline{\tenbf FIELD IDENTIFICATION IN}
\centerline{\tenbf NONUNITARY DIAGONAL COSETS}
\vglue 24pt
\centerline{\eightrm PIERRE MATHIEU and DAVID S\'EN\'ECHAL}
\centerline{\eightit D\'epartement de physique, Universit\'e Laval,
Qu\'ebec, Canada G1K 7P4. }
\vglue 10pt\centerline{\tenrm and}\vglue 10pt
\centerline{\eightrm MARK WALTON}
\baselineskip=10pt
\centerline{\eightit Physics Department, University of Lethbridge}
\centerline{\eightit Lethbridge (Alberta) Canada T1K 3M4.}
\baselineskip=10pt
\vglue 20pt


\vglue 16pt
\centerline{\eightrm ABSTRACT}
{\rightskip=1.5pc
\leftskip=1.5pc
\eightrm\parindent=1pc

We study the nonunitary diagonal cosets constructed from
admissible representations of \KM algebras at fractional
level, with an emphasis on the question of field identification.
Generic classes of field identifications are obtained from the analysis of
the modular $S$ matrix. These include the usual class related to outer
automorphisms, as well as some intrinsically nonunitary field
identifications.
They allow for a simple choice of coset field representatives where
all field components of the coset are associated with integrable
finite weights.

\vglue12pt}

\baselineskip=13pt
\immediate\openout\refs=references
\immediate\write\refs{\ninerm\baselineskip=11pt}
\section{Introduction}

Most unitary rational conformal field theories can be described in terms of
coset models. The properties of the conformal theories can thus be extracted
from those of the building blocks of the coset. Unitarity of the conformal
field theories is guaranteed by restricting the levels of \KM algebras
forming the cosets to the set of positive integers.

However, unitary solvable models in conformal field theory constitute
a small class among all solvable models.
The characteristic property of solvable models is modular invariance.
As observed recently [\ref{V. Ka\v c and M. Wakimoto, \jr{Proc. Nat. Acad.
Sci. USA}\vol{85} (1988) 4956.}-\aref{V. Ka\v c and M. Wakimoto,
\jr{Adv. Ser. Math. Phys.}\vol7 (World Scientific, 1988) 138.}\aref{V. Ka\v c
and M. Wakimoto, \jr{Branching functions for winding subalgebras and tensor
products}, MIT preprint.}\aref{P. Mathieu and M. Walton,
\jr{Prog. Theor. Phys. Supp.} \vol{102} (1990) 229.}\refname\Mathieu\aref{S.
Mukhi and S. Panda, \jr{Nucl. Phys.}\vol{B338}
(1990) 263.}\refname\Mukhi\aref{A. Kent, \jr{PhD thesis} Cambridge
University (1986).}\aref{H. Riggs, \jr{Nucl. Phys.} \vol{B326} (1989)
67.}\aref{T. Nakanishi, \jr{Nucl. Phys.}\vol{B334} (1990) 745.}\ref{A. Kuniba,
T. Nakanishi and J. Suzuki, {\it Ferro- and antiferromagnetization in
RSOS models}, Tokyo preprint (1990).}\refname\Kuniba],
nonunitary conformal models can also be described
by cosets. The price to pay for the non-unitarity is the presence in
the coset of \KM algebras at fractional level.
Nevertheless, even if the level  is fractional, there are {\it admissible}
representations which are covariant with respect to modular
transformations [1-3].
This is the root of modular covariance for the related conformal
field theories.

In this paper we will specifically study diagonal cosets of the form
$$ {\gh_m\oplus\gh_l\over\gh_{m+l}} \eqlabel\diagCos $$
where
$$ m = t/u\com l\in{\bf N}\quad.\eq $$
Here $\gh_m$ is some affine Lie algebra at level $m$
(whose finite restriction is denoted $\gb$).
The corresponding central charge is
$$ c = {l\dim\gb\over l+g}\l\{1 - {g(g+l)\over l^2}
{(p'-p)^2\over pp'}\r\}\eq$$
where we have introduced the integers $p$ and $p'$ defined by
$$m+g={lp\over (p'-p)}\com p'-p = lu\eqlabel\pprime$$
($g$ is the dual Coxeter number of $\gb$.)
For $l=1$, these cosets describe minimal models of $W\gb$
conformal algebras [\ref{V.A. Fateev and S. Lykyanov,
\jr{Int. J. Mod. Phys.}\vol{A3} (1988) 507; {\it Additional
symmetries in two-dimensional conformal quantum field theory and exactly
solvable models. I. Quantization of Hamiltonian structures. II. The
representations of W-algebras. III. Minimal models}, Kiev preprints
(1988).}\refname\Fateev].
For $l>1$ the chiral algebra is known only for very few cases.
However, when $\gb=su(N)$, those cosets are in the universality class
of the fused RSOS models introduced in [\Kuniba].
The most famous nonunitary coset model is the Yang-Lee
singularity [\ref{J. L. Cardy, \jr{Phys. Rev. Lett.}\vol{54}
(1985) 1354.}\refname\Cardy].

We will be concerned mainly with field identification
for diagonal nonunitary cosets, i.e. with the
problem of determining the classes of coset fields which share the same
characters, and which are thus undistinguishable.
Our main result will be that it is always possible to choose
representatives of those classes constructed solely from
weights, whose finite parts are integrable.
The discussion will be illustrated with many examples.

\section{Ka\v c-Moody algebras at fractional level.}

In this section we review the  results of Ka\v c and Wakimoto concerning
\KM algebras at fractional level. This is preceded by a brief review of
general results on \KM algebras and their relation to conformal theories
through WZNW models, which also fixes the notation.

\subsection{A brief review of \KM  algebras.}

We will only consider untwisted affine \KM algebras, that is those that are
the central extensions of loop algebras of finite Lie algebras.
The generators $J_{a,n}$ of the \KM algebra $\gh$ obey the following
commutation rules:
$$\big[ J_{a,n} , J_{b,p} \big]\ =\ i{f_{ab}}^c J_{c,n+p} +
mK_{ab}\delta_{n+p} \eqlabel\KMalgebra$$
where ${f_{ab}}^c$ are the structure constants of
a finite Lie algebra $\gb$ of rank $r$,
$K_{ab}$ is the corresponding Killing form, and $m$ is the {\it level}.
Each generator $J_{a,n}$ carries a Lie index $a$ and a Fourier mode index
$n\in\Zb$. The zero-modes $J_{a,0}$ generate a finite Lie algebra
$\gb\subset\gh$.
To give (\KMalgebra) properties similar to those of finite Lie algebras,
we introduce two additional generators $\mh$ and $\nh$ such that
$$ [\nh,J_{a,p}] = pJ_{a,p} \and [\mh,J_{a,p}] = [\mh,\nh] =0 \eq $$
The Cartan subalgebra is therefore generated by $\mh$, $\nh$ and $J_{h,0}$,
where the index $h$  refers to the Cartan sub-algebra of $\gb$.
Thus, the roots of $\gh$ have the following form:
$$ \al = (\alb,0,n) \eq $$
where $\alb$ is a root of $\gb$, or 0, and where $n$ is the grade of $\al$
in the root lattice of $\gh$. The 0 in the last entry simply
means that $\mh$ commutes with everything, and just gives the level when
applied to a state ($\mh$ was introduced in order to treat weights with
different levels within the same framework). The roots are at level 0,
and do not depend on the central extension of (\KMalgebra).
The $r+1$ simple roots are taken to be
$$\eqalign{
\al_i &= (\alb_i,0,0)\qquad i=1,\dots,r\cr
\al_0 &= (-\thb,0,1)\cr}\eq $$
where the $\alb_i$ are the simple roots of $\gb$ and
$\thb$ is the highest root of $\gb$.
The simple root $\al_0$ was chosen such as to make all roots
$(\alb,0,n)$ with $n>0$ positive.
Given two weights $\la=(\lab,m,n)$ and $\la'=(\lab',m',n')$, one
extends the inner product defined on the weight space of $\gb$ to the
following inner product:
$$ (\la,\la') := (\lab,\lab') + nm' + n'm \eq $$
This allows us to define for $\gh$ quantities taken from finite Lie
algebras.

Corresponding to a root $\al$, we define the coroot $\al^\vee~$:
$$\al^\vee~=~{{2\al}\over{(\al,\al)}}~=~r_\al\al\eqlabel\coroot$$
For long roots, $(\al,\al)=2$ by convention, and the
coroots then coincide with the roots ($r_\al=1$). For short roots (absent from
simply laced algebras) the coroots are integral multiples of the roots
($r_\al=2$ or 3).
The fundamental weights $\omega^\mu$ are defined as
$$ (\omega^\mu,\al_\nu^\vee)~=~\delta^\mu_\nu\eq$$
By convention, Greek indices will run from 0 to $r$ and pertain to
affine quantities, whereas Latin indices will run from 1 to $r$ and
pertain to the finite restrictions; thus we can identify $\al_i$ and
$\alb_i$.
Any affine weight $\la=(\lab,m,0)$ of grade zero
can be uniquely decomposed with respect to the basis $\{\omega^\mu\}$:
$$\la~=~\sum^r_{\mu=0}~\la_\mu \omega^\mu\com\eqlabel\affw $$
and may also be represented by the vector
$\la~=~[\la_0,\la_1,\ldots,\la_r] $ of its Dynkin labels.
For those weights, only the finite parts are important in inner
products: $(\la,\mu) = (\lab,\mub)$.
The finite Dynkin labels coincide with those of
$\lab$:  $\la_i=\lab_i$.
One also defines the marks and comarks $k_\mu$ and $k^\vee_\mu$ by
$$\thb~=~\sum_{i=1}^r\ k_i\al_i~=~\sum_{i=1}^r\ k^\vee_i\al^\vee_i\com\eq$$
with $k_0=k^\vee_0=1$ and $r_\mu=k_\mu/k^\vee_\mu$.
The marks and comarks are always positive integers.
For $\su{N}$ they are all equal to 1.
The level $m(\la)$ of a weight is
$$ m(\la)~=~\sum^r_{\mu=0}~\la_\mu k^\vee_\mu\eqlabel\LEV $$
Once the level $m$ of an affine weight $\la=(\lab,m,0)$ is fixed, there is a
one-to-one correspondence between affine and finite weights.

The Weyl group of $\gb$ is denoted $W$, and is
generated by the $r$ primitive (simple) reflections $s_i$
associated with the simple roots $\al_i$.
Their action on a weight $\la$ is $ s_i\la ~=~\la - \la_i~\al_i$.
Each element $w$ of $W$ has a parity $\epsilon(w)$ equal to 1 (-1) if
$w$ is expressible as an even (odd) number of primitive reflections.
The {\it shifted action} of $w\in  W$ is defined by
$$ w\bullet\la\ :=\ w(\la+\rho) - \rho\eq$$
with $\rho$ given  by
$$\rho\ =\ \sum^r_{\mu=0} \omega^\mu\ =\ [1,1,\ldots,1]\eqlabel\rhO $$
Appendix A briefly describes the roots, coroots and Weyl groups of the four
classical algebras.
For further details, the reader is referred to [\ref{V. Ka\v c,
\jr{Infinite-dimensional Lie algebras,} third ed. (Cambridge University
Press, Cambridge, 1990).},\ref{P. Goddard and D. Olive, \jr{Int. J. Mod.
Phys.}\vol{A1} (1986) 303.}].

\subsection{\KM algebras at integer level and WZNW models.}

A weight $\la$ is integral if all its Dynkin labels $\la_\mu$ are integers.
Further, if $\la_\mu\geq 0,$ then $\la$ is the highest weight
of an integrable representation of $\gh$.
These representations are unitary and can be integrated to
representations of the \KM group.
Since the comarks are positive integers, we see from (\LEV)
that the level of integrable highest weights is always integer.
The set of integrable highest weights at fixed level $m$ is denoted $P^m_+.$

The current algebra of a WZNW model
based on $\gb$ is exactly (\KMalgebra) for some positive integer
level [\ref{E. Witten, \jr{Commun. Math. Phys.}\vol{92} (1984) 455.}].
Each primary field of the WZNW model creates states filling out an integrable
highest weight representation of $\gh$ at level $m$ [\ref{V.G.
Knizhnik and A.B. Zamolodchikov, \jr{Nucl. Phys.}\vol{B247} (1984) 83.}].
Thus WZNW primary
fields and integrable highest weights are in one-to-one correspondence and the
latter may be used to label the primary fields.

The conformal weight of the primary field $\la$ is
$$ h_\la~=~{(\la,\la + 2\rho)\over 2(m+g)}~=~
{|\la +\rho |^2 -|\rho|^2\over 2(m+g)} \eqlabel\WEIGHT $$
where $\rho$ is given by (\rhO) and $g$ is the dual Coxeter number
of $\gb,$ the level of $\rho$:
$$ g\ =\ \sum^r_{\mu=0} k^\vee_\mu\eq$$

The Sugawara construction
$$ L_n~=~{1\over 2(m+g)}
\sum_{l\in\Zb} K^{ab} : J_{a,n+l} J_{b,-l}:~-~\delta_{n,0}
{c_{\gb}\over 24}\eqlabel\SUG
$$ associates with $\gh$ the Virasoro algebra
$$\big[L_n,L_p \big]~=~(n-p) L_{n+p}~+~
{c_{\gb}\over 12} n(n^2-1) \delta_{n+p,0}\eq $$
with the central charge
$$ c_{\gb} ~=~{{m \dim{\gb}}\over m+g}\eqlabel\C $$
The colons in (\SUG) indicate the standard normal ordering.

One defines the characters of the integrable representation as
$$ \chi_\la (\tau,z)~=~\Tr_\la
\exp \l\{ 2\pi i\tau\Big[ L_0 - {c\over {24}} + J_0\cdot z \Big] \r\}
\eq $$
where $J_0\cdot z = \sum_h z_hJ_{h,0}$.
The specialized characters are obtained by setting $z=0$:
$$ \chi_\la (\tau)~=~\Tr_\la \exp \bigg\{ 2\pi i\tau\Big( L_0 - {c\over{24}}
\Big) \bigg\}\eq $$
The characters are the building blocks of the partition functions of
WZNW models on the torus, and consequently have simple modular properties.
Under the modular transformation $S:\tau\mapsto-1/\tau$,
The characters transform amongst each other [\ref{V. Ka\v c and
D. Peterson, \jr{Adv. Math.}\vol{53} (1984) 125.}\refname\KacPeter]:
$$ \chi_\la(-1/\tau) = \sum_\mu~S_{\la\mu}~\chi_\mu(\tau) \com\eq $$
where
$$ S_{\la\mu}~=~F_m\sum_{w\in W}\ \ep(w)
\exp\bigg\{-{{2\pi i}\over {m+g}} \Big( w(\la +\rho), \mu +\rho\Big)\bigg\}
\eqlabel\matriceS $$
The constant of proportionality $F_m$ is independent of $\la,\mu$ and it can be
fixed by the unitarity of the matrix $S$:
$$ F_m\ =\ i^{|\bar\triangle_+|} |M^*/(m+g)M|^{-{1\over 2}}\eqlabel\FmA$$
where $|\bar\triangle_+|$ is the number of positive roots of $\gb,$ $M$ is the
lattice generated by the long roots of $\gb,$ and $M^*$ is its dual.
(For simply laced algebras, $M$ is just the root lattice.)
Under the transformation $\tau\mapsto\tau+1$ (the other generator of the
modular group) the characters simply acquire a phase factor:
$$ \chi_\la(\tau+1) = \exp\l\{2\pi i (h_\la - c/24)\r\}\chi_\la(\tau)
\eq $$

{}From the modular $S$ matrix one defines the charge conjugation
matrix ${\cal C}$ by
[\ref{E. Verlinde, \jr{Nucl. Phys.}\vol{B300} (1988) 389.}]
$$ {\cal C}S = S^* \qquad{\rm or}\qquad {\cal C} = S^2\eq $$
A direct calculation yields ${\cal C}_{\la,\la'} = \delta_{C\la,\la'}$, where
the integrable weight $C\la$ is the charge conjugate of $\la$,
given by [\ref{V. Ka\v c and M. Wakimoto,
\jr{Adv. Math.}\vol{70} (1988) 156.}]
$$ C\la = -w^0\bullet\lambda  \eq $$
$w^0$ being the longest element of $W$.
Charge conjugation is related to the reflection symmetry of the Dynkin diagram
of $\gb$. For $su(N)$, charge conjugation of an integrable weight amounts to
reversing the order of the finite Dynkin labels $\la_i$.

\subsection{Admissible representations of \KM algebras at fractional level.}

Consider a fractional level
$$ m\ =\ {t\over u}$$
where $t\in\Zb,\ u\in{\bf N}$ and $(t,u)=1.$ When $u\neq 1$ (fractional
level) representations of affine \KM algebras are necessarily nonunitary. But
some of these still have modular properties analogous to those of the
integer-level unitary integrable representations.
Ka\v c and Wakimoto [1-3] discovered a class of rational
level highest weight representations that are modular invariant and include the
integrable unitary representations. These they called admissible
representations. Their highest weights may be described as follows.

A fractional weight $\la$ at level $m$ will be admissible if it
satisfies the following two conditions:
$$\eqalign{
&i)\quad(\la+\rho,\al^\vee) \not\in -\Zb_+\com
\forall\al^\vee\in R_+ \cr
&ii)\quad{\bf Q} R^\la = {\bf Q} \Pi}\quad\eqlabel\admiA$$
where $\Pi:=\{\alv_0,\alv_1,\dots,\alv_r\}$ is the set of simple coroots,
$R_+$ is the set of positive real coroots of $\gh$, and where
$$R^\la := \{\alv| (\la,\alv) \in \Zb\}\qquad.\eqlabel\Rlambda$$
The second equation simply states that the rational span of the set $R^\la$
should be the same as the rational span of the simple coroots.
However, we will use a different and more constructive characterization of
admissible weights, also used by Ka\v c and Wakimoto [1],
whose compatibility with the above definition will be shown in appendix B:
To every element $y$ of the Weyl group $W$ is associated a set of
possible admissible highest weights $\la.$ Furthermore, each of these weights
may be broken up into an integer (I) and a fractional (F) part:
$$ \la~=~y\bullet \l( \la^I - (m+g)\la^{F,y} \r)\eqlabel\YLA $$
where $\la^I$ and $\la^{F,y}$ are both integral weights. The level of the
integer part $\la^I$ is
$$ m^I\ =\ u(m+g) - g\  \geq 0\eq $$
while that of the fractional part $\la^{F,y}$ is
$$ m^F\ =\ u - 1\ \geq 0\eq$$
In addition, the integer part $\la^I$ is the
highest weight of an integrable highest weight representation,
$$ \la^I\ \in\ P^{m^I}_+ \eq $$
The characterization of the fractional part is more complicated. The Dynkin
labels of $\la^{F,y}$ must satisfy
$$ \la^{F,y}_\mu~\in~r_\mu\Zb\eqlabel\LAFKZ $$
and be such that
$$ \ct \la^{F,y}_\mu + y\big(\al^\vee_\mu\big) \in R_+
\eqlabel\LAFY $$
where $\tilde c$ is
the canonical central element
$$ \ct~=~\sum_{\mu=0}^r\ k^\vee_\mu\al^\vee_\mu~=~ (0,1,0)\eq $$
Note that $k_\mu/k^\vee_\mu$ is always an integer, most often 1
(see appendix A).
Thus given $y\in W$ one can determine the possible values of
$\la^{F,y}_\mu$ at a given level $m^F$ and then construct the admissible
weights $\la$ of level $m$ corresponding to the choice of $y.$ This set of
admissible highest weights for a fixed $y$ will be denoted $P^m_y.$ The set of
all admissible highest weights at level $m$ is just the union of these:
$$ P^m\ =\ \bigcup_{y\in W}\ P^m_y\eq $$
When $u=1,$ we find $P^m=P^m_+.$

It turns out that not all $y\in W$ need be
considered. An admissible $\la$ may have more than one decomposition of the
form (\YLA), each one corresponding to a different $y$.
In fact, as shown in [\Mathieu] we can
restrict $y$ to a subgroup of the finite Weyl group
$W$, namely $W/W(A)$, where $W(A)$ is the subgroup of $W$
that is isomorphic to the outer automorphism group  $O(\gh)$ of $\gh$.
Recall that outer automorphisms permute the simple roots
of $\gh$ in ways that preserve the Dynkin diagram (or the Cartan matrix).
Their effect on weights is to permute the Dynkin labels.
To each $A\in O(\gh)$ there
corresponds an element $w_A$ of $W$ according to the relation [\ref{D.
Bernard, \jr{Nucl.Phys.}\vol{B288} (1987) 389.}]
$$ A(\la-m\omega^0) = w_A(\la-m\omega^0)\eqlabel\AWA $$
The outer automorphisms of specific algebras are described in
appendix A.

Finally let us mention some properties of the set of admissible weights
at fractional level.
Consider the proper fractional part of an admissible weight:
$$ \la^F ~:=~ y(\la^{F,y}) \eqlabel\fracF $$
These fractional parts satisfy an additivity property  modulo $u$, i.e.,
given two admissible fractional parts $\la^F$ and $\mu^F$, there exists
an admissible fractional part $\nu^F$ such that\foot{This property
is actually a conjecture, and has been checked on
many examples.}
$$ \lab^F + \mub^F ~=~ \bar\nu^F + u \bar\xi\com\eq$$
where $\bar\xi$ is an integer weight.
This means that the finite weights $u\lab^F$ belong to $\Gamma/(u\Gamma)$,
where $\Gamma$ is the lattice of integral finite weights.
The number of distinct fractional part is therefore $u^r$.
Since the number $N_i(m^I)$ of admissible integral parts $\la^I$ depends
only on the algebra and the integer level $m^I$, we conclude that the
number of admissible weights at level $m=t/u$ is
$$ N(m) ~=~ u^r~N_i(m^I) \eq $$
Notice that $N_i(0)=1$, and that, for $su(r+1)$,
$$ N_i(m^I)~=~{(m^I+r)!\over r!~m^I!}\eq $$
This additivity property modulo $u$ for fractional parts is useful
when considering fusion rules [\Mathieu].
Let us also mention the obvious fact that the set of admissible weights
at level $m$ is symmetric with respect to all outer and inner automorphisms
of the algebra $\gh$. For $\su3$ at $m^I=0$, this leads to a pattern of
admissible weights ressembling a star of David.
\subsection{Examples of admissible representations.}

To make the above discussion more specific, we will illustrate it for
$\su2$, $\su3$ and $\so5$.
In the first two cases all marks and comarks are 1, and hence
$\la^{F,y}_\mu\in\Zb$.
For $\su2$ we will use the following notation:
$$ \la^I~=~[m^I-n,~n]\com\la^{F,y}~=~[m^F-k,~k]
\eqlabel\SUTWO $$
where $n$ and $k$ are integers.
One  first fixes an element $y\in W/W(A)$ and derives the corresponding
restrictions on the values of $k$ from (\LAFY).
Since $W=W(A)$ in the case of $\su2$, the only necessary $y$ is the
identity. Then (\LAFY) reduces to the following two requirements:
$$\l\{\eqalign{
&(m^F-k+1)\al^\vee_0 + (m^F-k)\al^\vee_1 \in\ R_+\cr
&k\al^\vee_0 + (k+1)\al^\vee_1 \in\ R_+\cr}\r.\eq $$
The coefficients of the coroots must be greater than or equal to zero, with
at least one being positive. This forces $m^F-k\geq 0$ and $k\geq 0.$
Therefore the two Dynkin labels must be positive definite, and any $\su2$
admissible weight is of the form $\la=\la^I-(m+g)\la^F$ with
$\la^I\in P_+^{m^I}$ and $\la^F\in P_+^{m^F}$.

For $\su3$ the finite Weyl group is $\{1,s_1,s_2,s_1s_2,s_2s_1,s_1s_2s_1\}$.
The elements of $W(A)$ are $\{1,s_1s_2,s_2s_1\}$, corresponding respectively
to the outer automorphisms 1, $a$ and $a^2$, where $a$ is a cyclic permutation
of the affine Dynkin labels.
Therefore one can restrict $y$ to the set $\{1,s_1\}$.
For $y=1$, (\LAFY) yields $\la_\mu^{F,1}\geq 0$. On the other hand,
for $y=s_1$, the inequalities are $\la_{0,2}^{F,s_1}\geq 0$ and
$\la_1^{F,s_1}\geq 1$.
For the specific example of level $-\frac32$ ($u=2$, $m^F=1$ and $m^I=0$),
the allowed $\la^{F,y}$ are
$$\eqalign{
\la^{F,1}~:~& [1,0,0],\quad [0,1,0],\quad [0,0,1]\cr
\la^{F,s_1}~:~& [0,1,0].\cr}\eq $$
Therefore there are 4 admissible highest weights:
$$ [-\frac32,0,0]\com[0,-\frac32,0]\com[0,0,-\frac32]\com
[-\hf,-\hf,-\hf]\eq$$
where the first three are obtained with $y=1$ and the last one with $y=s_1$.

Let us now consider $\so5$.
The simple roots are $\al_0=[2,0,-2]$, $\al_1=[0,2,-2]$ and $\al_2=[-1,-1,2]$,
the last one being the short root. All marks and comarks are equal to 1
except $k_2=2$. This implies that $\la_{0,1}^{F,y}\in\Zb$ and
$\la_2^{F,y}\in 2\Zb$. The Weyl group is
$\{1,s_1,s_2,s_1s_2,s_2s_1,s_1s_2s_1,s_2s_1s_2,s_1s_2s_1s_2\}$
and its subgroup $W(A)$ is $\{1,s_1s_2s_1\}$.
Therefore it is sufficient to consider $y$ among the set
$\{1,s_1,s_2,s_1s_2\}$.
The constraints on $\la^{F,y}$ are easily found to be
$$\eqalign{
&\la^{F,1}_\mu\geq 0~~;\cr
&\la^{F,s_1}_{0,2}\geq 0~~;~~ \la^{F,s_1}_1\geq 1~~;\cr
&\la^{F,s_2}_{0,1}\geq 0~~;~~ \la^{F,s_2}_2\geq 1~~;\cr
&\la^{F,s_1s_2}_0\geq -1~~;~~\la^{F,s_1s_2}_1\geq 0~~;~~
\la^{F,s_1s_2}_2\geq 2~~;\cr}\eq $$
Notice that, in fact, $\la^{F,s_2}_2\geq 2$, and since the other two
Dynkin labels must be positive, the sector $y=s_2$ is allowed only for
$u\geq3$.

\subsection{The associated Weyl subgroup.}

Given an admissible weight $\la$, we may define the so-called
associated Weyl subgroup $W^\la\subset W$ [1-3].
It is spanned by the reflections with
respect to all positive roots $\alb$ such that $(\la,\alb^\vee)\in\Zb$.
If the finite Dynkin labels are integers, it coincides with the
full Weyl group.
On the other hand, if some of the finite Dynkin labels are not integers,
$W^\la$ will be a proper subgroup of $W$.
For instance, the associated Weyl subgroup corresponding to the $su(3)$
weight $(\hf,-\frac32)$ is $\{1,s_1s_2s_1=s_\theta\}$.
As another example, consider the $su(4)$ weights $(-\frac13,0,-\frac53)$
and $(-\frac13,-\frac23,-\frac53)$.
In the first case the two positive roots such that $(\la,\al^\vee)\in\Zb$ are
$\al_2$ and $\theta$, and thus $W^\la=\{1,s_2,s_1s_2s_3s_1s_2s_1\}$.
For the second case, the corresponding set of positive roots is
$\{\al_1+\al_2\}$ and $W^\la=\{1,s_1s_2s_1\}$.

Given an admissible weight $\la$, it is generically possible to find
elements $w$ of the Weyl group such that $w.\la$ is also admissible.
In fact, these elements belong to the coset $W/W^\la$.
In other words, for any $w\in W$, there is a
unique element $\wt\in W^\la$ such that $w\wt.\la$ is also admissible.
The proof is given in appendix D.
If $n(W)$ is the order of the group $W$, then admissible weights occur
in sets comprising $n(W)/n(W^\la)$ elements, related by shifted actions.
Within each set the conformal dimension is the same, since (\WEIGHT) is
invariant with respect to shifted actions of $W$.
If $\la$ and $w.\la$ are admissible weights, they both admit a
decomposition of the form (\YLA), with elements $y$ and $y'$ of W
respectively. In general $y$ and $y'$ are not unique, but we can
choose them such that $y'=wy$.
This assertion is also proven in Appendix D.
If we insist on picking $y$ and $y'$ from a fixed set of
representatives of $W/W(A)$, then the relation between $y$ and $y'$ is
rather $wy=y'w_A$, for some $w_A\in W(A)$.

\subsection{Modular properties of characters for admissible representations.}

Let us now discuss the modular transformation properties of the characters
of the highest weight admissible representations.
For the untwisted \KM algebras, Ka\v c and Wakimoto [1-3] found that the
characters transformed according to the following modular
$S$ matrix:\foot{In the case of non-simply laced algebras, this formula
is only valid when $u$ is not a multiple of the ratio $r_\mu$ corresponding
to the short roots. Thus, this formula should not be used when
$u$ is even for $B_r$, $C_r$ and $F_4$, and when $u\in 3\Zb$ for $G_2$:
The corresponding admissible sets are ill-defined from the point of view
of modular transformations.}
$$\eqalign{
S_{\la\mu }&= F_m\ep(yy')
\exp \Big\{ 2\pi i\big[ (\la^I+\rho,\mu^F)+(\la^F,\mu^I+\rho)-(m+g)
(\la^F,\mu^F)\big]\Big\}\cr
&\qquad\times \sum_{w\in W}~\ep(w)\exp\Big\{{{-2\pi i}\over{m+g}}
\big( w(\la^I+\rho), \mu^I+\rho \big)\Big\}\cr}\eq $$
where
$$F_m~=~i^{|{\bar\triangle}_+|}
\l| {{M^*}\over{u^2(m+g)M}}\r|^{-{1\over 2}}\eq $$
Here $\la\in P^m_y$ and $\mu\in P^m_{y'}$
(the fractional parts $\la^F$ and $\mu^F$ are defined as in (\fracF)).
This matrix is {\it unitary} (this corrects statements to the contrary made in
[\Mathieu]).

Notice that the summand depends only on the integer part $\la^I$.
This almost immediately implies that the fusion rules are essentially
determined by the integer part (up to sign , in fact. e.g. see [\Mathieu]).
When $m^F=0,$ we only have to consider $y=1,$ and the modular matrix $S$
reduces to the one found by Ka\v c and Peterson [\KacPeter] in the integrable
case (see Eq.(\matriceS)).

For manipulations of the phase factors in front of the summand, the
following result is useful [\Mathieu]:
If $\zeta$ and $\nu$ are integrable weights, then
for any $w\in W$ we have
$$\big( (w-1)\zeta, \nu\big)~=~0~\mod1\com\eqlabel\MODLEM$$
provided
$$\zeta_\mu\in r_\mu\Zb\quad{\rm
or}\quad\nu_\mu\in r_\mu\Zb\com\eqlabel\EITHER$$
a condition satisfied by the fractional parts of admissible weights.
With this result, the modular $S$ matrix may be rewritten in the
following form:
$$ S_{\la\mu} = F_m \ep(y)
\exp \Big\{ 2\pi i\big[ (\la^F,\mu + \rho)\big]\Big\}
\sum_{w\in W}\ep(w)\exp\Big\{{{-2\pi i}\over{m+g}}
\big(w(\la^I+\rho), \mu + \rho \big)\Big\}.\eqlabel\S $$

As an example, let us consider the $\su2$ modular matrix.
For $\su2$ we only need to consider $y=1.$ In the notation
(\SUTWO), the modular matrix $S$ is easily seen to be
$$ S_{\la\la'}~=~\l[{2\over{u^2(m+2)}}\r]^\hf
(-1)^{k(n'+1)+k'(n+1)}e^{-\pi imkk'}
\sin\l[{\pi(n+1)(n'+1)\over m+2}\r] \eqlabel\Ssutwo $$
Other examples are presented in the following sections.

Notice that when the level is fractional there are admissible
representations with negative conformal weights, in contrast with
unitary models, where the identity primary field has the lowest conformal
weight (zero).
In other words, the conformal
weight formula (\WEIGHT) is not positive semi-definite, and the identity is
no longer the true vacuum. Of course, since the number of representations
is finite, there still exists
a field (in general not unique) with minimum conformal weight.

\subsection{Charge conjugation.}

Let us now turn to a description of charge conjugation for weights at
fractional level.
The charge conjugation matrix ${\cal C}$ was calculated by Ka\v c and Wakimoto
from the general formula ${\cal C}=S^2$.
Let us denote by $w^\la$ the longest element of $W^\la$, the Weyl
subgroup associated with $\la$. ${\cal C}$ was found to be [3]
$$ {\cal C}_{\la,\la'} = \ep(w^0)\ep(w^\la) \delta_{C\la,\la'}\eq $$
where
$$C\la := -w^\la\bullet\lambda\quad.\eqlabel\Cwlambda$$
Obviously charge conjugation in the fractional case is not related to
inner automorphisms of the algebra.
For $\su2$, it takes a particularly simple form.
In that case, $\lab=n-(m+2)k$, which is always fractional if $k\not=0$.
If $k=0$, $W^\la=W=\{1,s_1\}$.
But $-s_1\bullet\la=\la$. Therefore, $C$ is the identity in the $k=0$ sector.
However, if $k\not=0$,  $W^\la=\{1\}$ and $-1\bullet\la=-\la-2\rho$,
so that
$$ \overline{C\la} = -n + (m+2)k -2 = m^I-n-(m+2)(u-k) \eq$$
Since $\ep(w^0)\ep(w^\la) = -1$, all entries in the $k\not=0$ sector
are $-1$'s. This result is easily checked directly from the expression
(\Ssutwo) for the $\su2$ modular S matrix.
The fact that for $\su2_m$ all weights with $k\not =0$ come as conjugate
pairs with the same conformal dimensions was noticed also in
[\Mukhi,\ref{J.D. Cohn, \jr{Phys. Lett.}\vol{B226} (1989) 267.}].
(From (\Cwlambda), it is obvious that $|C\la+\rho|^2=|\la+\rho|^2$,
which implies that $h_{C\la} = h_\lambda$).

Let us give another example of eq.(\Cwlambda) for $su(4)$.
The longest element of $W$ is $w^0=s_1s_2s_3s_1s_2s_1$, with $\ep(w^0)=1$.
The element $w^\la$ corresponding to $\la=(-\frac13,0,-\frac53)$ is also
$w^0$ so that
$$ C(-\frac13,0,-\frac53)
{}~=~-w^0\bullet(-\frac13,0,-\frac53) = (-\frac53,0,-\frac13) \eq $$
Similarly, for the weight $(-\frac13,-\frac23,-\frac53)$, $w^\la=s_1s_2s_1$:
$\ep(w^0)\ep(w^\la)=-1$ and
$$  C(-\frac13,-\frac23,-\frac53)
{}~=~-s_1s_2s_1\bullet(-\frac13,-\frac23,-\frac53) =
(-\frac13,-\frac43,-\frac13) \eq $$

\vglue 12pt
\centerline{Table 1 : $\su3$ at $m=-\frac94$.}\nobreak
$$\vbox{\eightrm\halign{
\hfil#\hfil&\qquad\hfil#\hfil&\qquad\hfil#\hfil&\qquad\hfil#\hfil&\qquad
\hfil#\hfil&\qquad\hfil#\hfil\cr
\multispan6\hrulefill\cr
label & $~~\la~~$ & $~~\la^{F,y}~~$ & $~~y~~$ & $~~w^\la~~$ & $~~C~~$\cr
\multispan6\hrulefill\cr
1 & $[-\frac94,0,0]$ & $[3,0,0]$ & 1 &$s_1s_2s_1$& 1\cr
2 & $[-\frac32,0,-\frac34]$ & $[2,0,1]$ & 1 & $s_1$ & 4 \cr
3 & $[-\frac34,0,-\frac32]$ & $[1,0,2]$ & 1 & $s_1$ & 3 \cr
4 & $[0,0,-\frac94]$ & $[0,0,3]$ & 1 & $s_1$ & 2 \cr
5 & $[-\frac32,-\frac34,0]$ & $[2,1,0]$ & 1 & $s_2$ & 7 \cr
6 & $[-\frac34,-\frac32,0]$ & $[1,2,0]$ & 1 & $s_2$ & 6 \cr
7 & $[0,-\frac94,0]$ & $[0,3,0]$ & 1 & $s_2$ & 5 \cr
8 & $[-\frac54,\frac14,-\frac54]$ & $[2,0,1]$ & $s_2$ &$s_1s_2s_1$& 13 \cr
9 & $[-\frac54,-\frac12,-\frac12]$ & $[1,0,2]$ & $s_2$ &$s_1s_2s_1$& 9 \cr
10 & $[-\frac12,-\frac12,-\frac54]$ & $[1,1,1]$ & $s_2$ &1& -16 \cr
11 & $[-\frac34,-\frac34,-\frac34]$ & $[1,1,1]$ & 1 &1& -15 \cr
12 & $[0,-\frac34,-\frac32]$ & $[0,1,2]$ & 1 &1& -14 \cr
13 & $[-\frac54,-\frac54,1]$ & $[0,0,3]$ & $s_2$ &$s_2$& 8 \cr
14 & $[-\frac12,-\frac54,-\frac12]$ & $[0,1,2]$ & $s_2$ &1& -12 \cr
15 & $[\frac14,-\frac54,-\frac54]$ & $[0,2,1]$ & $s_2$ &1& -11 \cr
16 & $[0,-\frac32,-\frac34]$ & $[0,2,1]$ & 1 &1& -10 \cr
\multispan6\hrulefill\cr}}$$
\vglue 12pt
\subsection{An example of `WZNW' model at fractional level.}

To conclude this section, we illustrate all previous results with a complete
example of `WZNW model' at fractional level: $\su3$ at level $-\frac94$.
The list of all admissible weights is given in table 1, together with the
corresponding values of $\la^I$, $\la^{F,y}$ and $y$, and the label of
the charge conjugated weight.

\def\h{\hfill}
The modular S matrix for this admissible set is ($\times 4$):
\vglue12pt
$$\left(\matrix{
\h 1&\h 1&\h 1&\h 1&\h 1&\h 1&\h 1&\h-1&\h
-1&\h 1&\h 1&\h 1&\h-1&\h-1&\h-1&\h 1\cr
\h 1&\h-1&\h 1&\h-1&\h-i&\h-1&\h i&\h-i&\h
 1&\h-1&\h i&\h-i&\h i&\h-i&\h i&\h 1\cr
\h 1&\h 1&\h 1&\h 1&\h-1&\h 1&\h-1&\h 1&\h
-1&\h-1&\h-1&\h-1&\h 1&\h 1&\h 1&\h 1\cr
\h 1&\h-1&\h 1&\h-1&\h i&\h-1&\h-i&\h i&\h
 1&\h-1&\h-i&\h i&\h-i&\h i&\h-i&\h 1\cr
\h 1&\h-i&\h-1&\h i&\h-1&\h 1&\h-1&\h i&\h
 1&\h-i&\h i&\h 1&\h-i&\h-1&\h i&\h-i\cr
\h 1&\h-1&\h 1&\h-1&\h 1&\h 1&\h 1&\h 1&\h
-1&\h 1&\h-1&\h 1&\h 1&\h-1&\h 1&\h-1\cr
\h 1&\h i&\h-1&\h-i&\h-1&\h 1&\h-1&\h-i&\h
 1&\h i&\h-i&\h 1&\h i&\h-1&\h-i&\h i\cr
\h-1&\h-i&\h 1&\h i&\h i&\h 1&\h-i&\h-1&\h
 1&\h i&\h-1&\h-i&\h-1&\h-i&\h 1&\h i\cr
\h-1&\h 1&\h-1&\h 1&\h 1&\h-1&\h 1&\h 1&\h
 1&\h-1&\h-1&\h 1&\h 1&\h-1&\h 1&\h 1\cr
\h 1&\h-1&\h-1&\h-1&\h-i&\h 1&\h i&\h i&\h
-1&\h-1&\h-i&\h-i&\h-i&\h-i&\h-i&\h 1\cr
\h 1&\h i&\h-1&\h-i&\h i&\h-1&\h-i&\h-1&\h
-1&\h-i&\h-1&\h-i&\h-1&\h-i&\h 1&\h-i\cr
\h 1&\h-i&\h-1&\h i&\h 1&\h 1&\h 1&\h-i&\h
 1&\h-i&\h-i&\h-1&\h i&\h 1&\h-i&\h-i\cr
\h-1&\h i&\h 1&\h-i&\h-i&\h 1&\h i&\h-1&\
 1&\h-i&\h-1&\h-1&\h-1&\h i&\h 1&\h-i\cr
\h-1&\h-i&\h 1&\h i&\h-1&\h-1&\h-1&\h-i&\h
-1&\h-i&\h-i&\h 1&\h i&\h-1&\h-i&\h-i\cr
\h-1&\h i&\h 1&\h-i&\h i&\h 1&\h-i&\h 1&\h
 1&\h-i&\h 1&\h-i&\h 1&\h-i&\h-1&\h-i\cr
\h 1&\h 1&\h 1&\h 1&\h-i&\h-1&\h i&\h i&\h
 1&\h 1&\h-i&\h-i&\h-i&\h-i&\h-i&\h-1\cr
}\right)$$

\section{Generalities on coset models}

Apart from WZNW models, which are unitary, most rational conformal field
theories can be described in terms of a coset model. The fields of such
models are associated with representations of the coset $\gh/\hh$, where
$\hh$ is a subalgebra of $\gh$. The crucial observation of GKO
[\ref{P. Goddard, A. Kent and D. Olive, \jr{Commun. Math. Phys.}
\vol{103} (1986) 105.}]
is that the energy-momentum tensor $T^{(\gh/\hh)}$ of the coset model is
simply the difference $T^{(\gh)}-T^{(\hh)}$ of the WZNW energy-momentum
tensors. The Virasoro modes are then simple differences:
$$ L_n^{(\gh/\hh)} = L_n^{(\gh)} - L_n^{(\hh)} \eqlabel\COSVIR $$
where $L_n^{(\gh)}$ and $L_n^{(\hh)}$  are given by the Sugawara construction
(\SUG). This implies that the coset central charge is
$$c_{\gh/\hh} = c_{\gh} - c_{\hh} \eq $$
where $c_{\gh}$ is given by (\C).

The primary fields of the coset are labelled by the admissible (or integrable,
if the level is an integer) representations of $\gh$ and $\hh$.
Let $\La$ and $\la$ be admissible weights of $\gh$ and $\hh$ respectively.
We will denote coset primary fields (or coset fields, for short)
by $\{\La;\la\}$.
Not all combinations of $\La$ and $\la$ are allowed. There are selection rules
which reflect the existence of relations between the centers of the
covering groups of $\gb$ and $\hb$ [\ref{G. Moore and Seiberg,
\jr{Phys. Lett.}\vol{B220}
(1989) 422.}\refname\Moore -\aref{D. Gepner, \jr{Phys. Lett.}
\vol{B222} (1989) 207.}\refname\Gepner\aref{W. Lerche, C. Vafa and N. Warner,
\jr{Nucl. Phys.}\vol{B324} (1989) 427.}\ref{C. Ahn and M. Walton,
\jr{Phys. Rev.}\vol{D41} (1990) 2558.}\refname\Ahn].
The allowed combinations of $\La$ and $\la$ are in fact specified by the
branching functions $b_\La^\la$ which tell how the character $\chi_\La$ can be
decomposed into characters $\chi_\la$ for admissible $\la\in\hh$:
$$ \chi_\La = \sum_\la~\chi_\la~b_\La^\la \eq $$
Generically
the branching functions $b_\La^\la$ are the characters of primary
fields in the coset models
(conditions under which this identification fails will be discussed below).
The selection rules imply a restriction on the above summation.

Although the generic relation between the characters of the coset primary
fields and those of the coset building blocks is not a simple ratio,
it turns out to be so for the modular matrices $S$ and $T$:
$$ S^{(\gh/\hh)}_{\{\La;\la\}\{\La';\la'\}} =
S^{(\gh)}_{\La\La'}~S^{(\hh)*}_{\la\la'} \eqlabel\cosetS $$
$$ T^{(\gh/\hh)}_{\{\La;\la\}\{\La';\la'\}} =
T^{(\gh)}_{\La\La'}~T^{(\hh)*}_{\la\la'} \eqlabel\cosetT $$
(Because of the unitarity and symmetry of $S$ and $T$, we replaced the
inverses of $S$ and $T$ by complex conjugates.)

A direct consequence of (\COSVIR) is that
$$ h_{\{\La;\la\}} = h_{\La} - h_{\la}~\mod1\quad,\eq $$
since $h$ is the eigenvalue of $L_0$.
(The reason for the $\mod 1$ is that coset primary fields may be built
from descendent WZNW fields).
This implies that the phase factor acquired by the character
$\chi_{\{\La;\la\}}$ under the modular transformation $\tau\mapsto\tau+1$ is
$$ \exp\l\{ 2\pi i\l( h_{\{\La;\la\}} -\frac1{24}c_{\gh/\hh}\r)\r\} \eq $$

Let us now discuss in more detail the selection rules
which prevent some combinations of weights $\La$ and $\la$ from being
coset primary fields, and the related question of field identification.
Here we will restrict ourselves
to the unitary case, leaving to the next section a detailed analysis of
the complications brought by non-unitarity.

As mentioned above, the selection rules reflect the existence of certain
relations between the centers of the covering groups of $\gb$ and $\hb$.
Denote by $G$ and $H$ the respective covering groups of $\gb$ and $\hb$,
and by $B(G)$ and $B(H)$ their centers.
There is an isomorphism between $B(G)$ and the group $O(\gh)$
of outer isomorphisms of $\gh$: To $A\in O(\gh)$ corresponds an element $\al$
of $B(G)$ whose eigenvalue on $\bar\La$, the highest weight of a $\gb$
representation, is $\exp[2\pi i(A\omega^0,\bar\La)]$.
This action extends uniquely to the affine case, with the result
$$ \al\La = \La e^{2\pi i(A\omega^0,\bar\La)} \eq $$
Similarly, let $\tilde\al\in B(H)$ be related to $\tilde A\in O(\hh)$.
Then the elements $\al$ and $\tilde\al$ of the centers can be
identified if and only if
$$ (A\omega^0,\La) = (\tilde A\omega^0,\la)~\mod1\eqlabel\centers $$
If this is not satisfied, the coset primary field $\{\La;\la\}$
does not appear.

Let us now turn to field identifications. The isomorphism between
$B(G)$ and $O(\gh)$ and the above selection rule imply that the
characters of $\{\La;\la\}$ and $\{A\La;\tilde A\la\}$ are equal.
At the level of the $S$ matrix, this implies [\Moore-\Ahn]
$$ S_{\{\La;\la\}\{\La';\la'\}} = S_{\{A\La;\tilde A\la\}\{\La';\la'\}}\eq$$
for all fields $\{\La';\la'\}$. Therefore the fields $\{\La;\la\}$ and
$\{A\La;\tilde A\la\}$ are not distinct, and should be
identified.

If there are no fixed points, i.e. no coset field $\{\La;\la\}$ and no
$A\in O(\gh)$ such that $\{\La;\la\} = \{A\La;A\la\}$ (strict equality),
then the string of field identification has a constant length.
For such cases, the branching functions are the characters of the coset
primary fields. If there are fixed points this is no longer
true.\foot{In the event of fixed points, the formula (\cosetS) yields
an $S^\dagger S$ different from the identity, but only in
some of its diagonal elements which are then
positive integers different from 1. The same is true of ${\cal C}=S^2$.}
The fixed points have to be resolved (for a general discussion and many
examples of fixed point resolution, see [\ref{A.N. Schellekens and
S. Yankielowicz, \jr{Nucl. Phys.}\vol{B334} (1990) 67.},\ref{A.N. Schellekens
and S. Yankielowicz, \jr{Int. J. Mod. Phys.}\vol{A5} (1990) 2903.}]).
The modular matrix as given in (\cosetS) always
describes how the branching functions transform under modular transformations.
Happily, fixed points do not prevent us from identifying fields using this
matrix.

\section{Nonunitary diagonal cosets.}

Let us now turn to diagonal cosets of the type (\diagCos) with $m$
fractional.
In the following we will denote a diagonal coset primary
field by $\{\gamma,\la;\la'\}$,
where $\gamma\in P_+^l$, $\la\in P_y^m$ and $\la'\in P_{y'}^{m+l}$.

Fixed points are ignored throughout.
Hence we freely identify branching functions with characters of the
coset fields. This is certainly not true when there are fixed points.
However, as far as field identification is concerned, this is
immaterial.

\subsection{Character decomposition.}

As already mentioned, not all combinations of admissible weights produce
coset primary fields. Allowed triplets are specified by the tensor product
decomposition of the characters:
$$\chi^{(l)}_\gamma \chi^{(m)}_\la = \sum_{\la'}
\chi_{\{\gamma,\la;\la'\}}~\chi^{(m+l)}_{\la'} \eqlabel\decomp $$
with the following conditions on $\la'$:
$$ \gamma+\la-\la' \in Q\com \la'\in P_y^{m+l}\com
\la^F=\la'^F\quad, \eqlabel\branchA $$
where $Q$ is the root lattice of $\gh$.
Notice that for a diagonal coset, (\centers) becomes
$(A\omega^0,\gamma+\la-\la')\in\Zb$, $\forall A\in O(\gh)$,
which is satisfied when
$\gamma+\la-\la'$ lies in the root lattice.
However, if $\la^F=\la'^F$, then $\gamma+\la-\la'$ is an integrable weight,
and the condition $\gamma+\la-\la'\in Q$ is equivalent to (\centers).
$\chi_{\{\gamma,\la;\la'\}}$ is the character of the primary field
$\{\gamma,\la;\la'\}$ (at this stage we ignore the question of fixed points).
Eq. (\decomp) has been proven by
Ka\v c and Wakimoto for $l=1$, and it is a conjecture of these authors in the
general case [1-3].
(It was also established  in [\Kuniba] for $\gh=su(N)$, under the
hypothesis that one can restrict oneself to the $y=1$ sector and ignore the
fractional part. The validity of these two assumptions is justified a
posteriori by our analysis of field identifications.)

Let us stress some of the salient features of the decomposition (\decomp):
\item{i)} $\la$ and $\la'$ are associated with the same Weyl group element
$y$, i.e., $y'=y$.
\item{ii)} The fractional parts of $\la$ and $\la'$ are equal. This
identification is made possible because $\la^F$ and $\la'^F$ have the same
level ($u-1$), despite the fact that the levels of $\la$ and $\la'$ are
different. The fractional part appears then as a conserved charge under
tensor product decomposition.
\item{iii)} The condition constraining the sum (\decomp) can be rewritten as
$$ \gamma + \la^I + l \la^{F,y} = \la'^I~\mod{Q} \eqlabel\branchB $$
Indeed, since $\la^{F,y} = \la'^{F,y}$, one has
$$ \gamma + \la - \la' = \gamma + y(\la^I - \la'^I + \la^{F,y}) \eq $$
The invariance of $Q$ under the action of $y$ yields (\branchB) from
(\branchA).

The dimensions of the coset primary fields can be extracted
from the expansion of the branching function in powers of $q=e^{2\pi i\tau}$.
However, since the latter is not known in general, we can only write
the fractional part of $h$, since
$$ h = h_\gamma + h_\la -h_{\la'}~\mod1\eqlabel\dimmod $$
This can be written in the form
$$ h = {|p(\la'+\rho)-p'(\la+\rho)|^2 - (p-p')^2|\rho|^2 \over
2lpp'} + {(\gamma,\gamma+2\rho)\over 2(l+g)} -{|\la-\la'|^2\over 2 l}
{}~\mod1\eqlabel\dimCos $$
in terms of the coprime numbers $p$ and $p'$ introduced in (\pprime).
Notice that
$$ p = m^I + g \com p' = m^I + g + lu \com\eqlabel\pprimeB $$
which fixes the bounds $p\geq g$ and $p'\geq g + lu$.
Notice that in terms of the integral and fractional parts of the
weights, (\dimCos) can be rewritten as
$$\eqalign{
 h = &{|p(\la'^I+\rho)-p'(\la^I+\rho)|^2 - (p-p')^2|\rho|^2 \over
2lpp'}\cr &\qquad + {(\gamma,\gamma+2\rho)\over 2(l+g)} -{|\la^I-\la'^I+
l\la^{F,y}|^2\over 2 l}~\mod1\quad,\cr}\eqlabel\dimCosI$$

For $l=1$ and simply laced algebras, the last two terms in  (\dimCos) cancel.
This can be seen as follows. One first rewrites the last two term under the
form (with $\la'-\la=\gamma-\beta$ for some $\beta\in Q$)
$$ {|\gamma+\rho|^2-|\rho|^2\over 2(g+l)} - {|\gamma|^2\over 2l}
+{2(\gamma,\beta)\over 2l} - {|\beta|^2\over 2l}\eq$$
For simply laced algebras $(\gamma,\beta)\in \Zb$ and $|\beta|^2\in 2\Zb$.
Therefore, modulo 1, the last two terms of the above disappear when $l=1$.
On the other hand, a characteristic feature of simply laced algebras
is that any $\gamma\in P^1_+$ can be written as $A\omega^0$ for some $A$.
Now, using (\AWA), one finds that
$$ {|A\omega^0+\rho|^2-|\rho|^2\over 2(g+1)} - {|A\omega^0|^2\over 2} =
(A\omega^0,w_A\rho) + {g\over 2}|A\omega^0|^2 \eq $$
Now, since $w_A\rho=A\rho-g(A\omega^0-\omega^0)$, this becomes
$$ -{g\over 2}|A\omega^0|^2 + (A\omega^0,\rho) \eqlabel\omeg $$
But it is true in general (i.e., for all Lie algebras) that the above
expression is an integer.

\subsection{Modular $S$ matrix for the coset.}

In terms of the coset components, the coset modular $S$ matrix is
$$ S_{\{\gamma,\la;\la'\}\{\xi,\mu;\mu'\}} =
S^{(l)}_{\gamma\xi} S^{(m)}_{\lambda\mu} \l[S^{(m+l)}\r]^*_{\la'\mu'}
\eqlabel\ScosetA $$
Up to a numerical factor which can be recovered from unitarity, the r.h.s. is
$$\eqalign{
S^{(l)}_{\gamma\xi}~&e^{2\pi i[(\la^I-\la'^I,\mu^F)+(\mu^I-\mu'^I,\la^F)
+ (\la^F,\mu^F)]}\qquad\times\cr
& \sum_{w\in W} \ep(w)e^{-2\pi i(w(\la^I+\rho),\mu^I+\rho)/(m+g)}~
\sum_{w'\in W} \ep(w')e^{2\pi i(w'(\la'^I+\rho),\mu'^I+\rho)/(m+g+l)}\cr}
\eqlabel\ScosetB$$
Since $\la^{F,y}_\mu\in r_\mu\Zb$, all the factors of $y$ in
(\ScosetA) can be eliminated, i.e., we can write $\la^{F,y}$ instead of
$\la^F$. Next, using the fact that
$(\nu+\beta,\sigma) = (\nu,\sigma)~\mod1$
if $\beta\in Q$ and if $\sigma_\mu\in r_\mu\Zb~\forall\mu$, one can
eliminate $\la^{F,y}$ and $\mu^{F,y}$ by means of (\pprimeB), i.e.
$$\eqalign{
\la^{F,y} &= \la'^I - \la^I - \gamma~\mod{Q} \cr
\mu^{F,y} &= \mu'^I - \mu^I - \xi~\mod{Q} \cr}\eq $$
Then (\ScosetB) can be expressed as
$$ S^{(l)}_{\gamma\xi}~e^{2\pi i(\gamma,\xi)}~
e^{-2\pi
i(\la^I-\la'^I,\mu^I-\mu'^I)}~\phi^{(m+g)}_{\la^I,\mu^I}~
\phi^{(m+g+l)*}_{\la'^I,\mu'^I}\eq$$
where we have defined
$$\phi^{(k)}_{\la\mu} := \sum_{w\in W} \ep(w)
e^{-2\pi i(w(\la+\rho), \mu+\rho)/k} \eq $$
For $l=1$ and the simply-laced algebras $A_r$ and $D_r$,
this result can be further simplified. Indeed, in that case
$$ S^{(1)}_{\gamma\xi}e^{2\pi i(\gamma,\xi)} = F_1\com \eq $$
where $F_1$ is defined in (\FmA).
The clue to the proof of this identity is again the observation that,
for those algebras, any level one
integrable weight can be written as $A\omega^0$ for a suitable outer
automorphism $A$.
Thus, writing $\gamma = A\omega^0$ and $\xi= A'\omega^0$, and using twice
the result
$$ S_{A\la^I,\mu^I} = e^{-2\pi i(A\omega^0,\mu^I)} S_{\la^I\mu^I}\com\eq $$
one easily finds
$$ S^{(1)}_{\gamma\xi}e^{2\pi i(\gamma,\xi)} =
e^{2\pi i(A\omega^0,A'\omega^0)} S^{(1)}_{A\omega^0,A'\omega^0} =
S^{(1)}_{\omega^0,\omega^0} = F_1 \eq $$
Hence, up to a multiplicative constant, the modular $S$ matrix for $A_r$ and
$D_r$ cosets at $l=1$ is
$$ S_{\{\gamma,\la;\la'\}\{\xi,\mu;\mu'\}} \propto
e^{2\pi i[(\la^I+\rho,\mu'^I+\rho)+(\la'^I+\rho,\mu^I+\rho)]}
{}~\phi^{(p/p')}_{\la^I,\mu^I}~\phi^{(p'/p)}_{\la'^I,\mu'^I} \eqlabel\ScosetC
$$
This reproduces the result found by Fateev and Lykyanov for the minimal
models of $Wsu(N)$ and $Wso(2N)$ conformal algebras, obtained
by means of the Feigin-Fuchs representation [\Fateev].
The expression (\ScosetC) is independent of $y$ and depends on
$\gamma$ and $\la^{F,y}$ only through the condition (\branchB) relating
$\la'^I$ to $\la^I$.
For simply laced algebras, for any choice of $\la^I$ there exists a $\gamma$
and a $\la^{F,y}$ yielding any possible  $\la'^I\in P^{m^I+u}_+$. Hence
one can equivalently describe a coset field by a doublet $(\la^I\mid\la'^I)$
where the two integrable weights are independent, the natural standpoint
in the Feigin-Fuchs description of $W\gb$ minimal models.

\subsection{Field Identification.}

Two coset primary fields $\phi$ and $\phi'$ can be identified if their
modular properties are identical, i.e. if, for any
field $\psi$, we have [\Gepner]
$$ S_{\phi\psi} = S_{\phi'\psi}\com \eqlabel\Sid $$
and if their conformal dimensions are the same.
This guarantees that the characters $\chi_\phi(\tau,z)$ and
$\chi_{\phi'}(\tau,z)$ are identical.
Let us now describe generic classes of field identifications.

\subsubsection{Field identification from outer automorphisms.}

As in the unitary case one expects the action of any element $A$ of the group
of outer automorphisms of $\gh$ to yield a field identification. Define the
action of $A$ on the coset field by
$$ A\{\gamma,\la;\la'\} = \{A\gamma,A\la;A\la'\}\eq $$
A straightforward calculation shows that
$$ AS^{(m)}_{\la,\mu} = S^{(m)}_{A\la,\mu} =
S^{(m)}_{\la,\mu}~e^{-2\pi i(A\omega^0,\mu)}~
e^{2\pi i((w_A-1)\la^F,\mu+\rho)} \quad.\eq $$
Therefore, for the coset $S$ matrix given by (\ScosetA), one has
$$ AS_{\{\gamma,\la;\la'\}\{\xi,\mu;\mu'\}} =
S_{A\{\gamma,\la;\la'\}\{\xi,\mu;\mu'\}} =
\al S_{\{\gamma,\la;\la'\}\{\xi,\mu;\mu'\}}\com \eq $$
where
$$ \al = e^{-2\pi i[(A\omega^0,\xi+\mu-\mu') + ((w_A-1)\la^F,\mu+\rho)
-((w_A-1)\la'^F,\mu'+\rho)]}\quad.\eqlabel\phaseA $$
Since $\la^F = \la'^F$, the last two terms in the phase become
$((w_A-1)\la^F,\mu-\mu')$. Furthermore, the branching rules demand that
$\mu-\mu'$ be an integral weight, in which case
$((w_A-1)\la^F,\mu-\mu')=0\mod1$. Similarly, $\xi+\mu-\mu'$ is an
element of the $\gb$ root lattice, whose inner product with $A\omega^0$
is necessarily an integer. Therefore $\al=1$ and $A$ indeed produces a
field identification.

It is also straightforward to check that the action of $A$ on coset
fields does not affect (\dimCos).
Indeed, from (\WEIGHT) and (\AWA), one has
$$ h_{A\la} = h_\la + (A\omega^0,w_A(\la+\rho)) +
\frac12(m+g)|A\omega^0|^2\eq$$
so that
$$ h_{A\gamma}+h_{A\la}-h_{A\la'} =
h_\gamma+h_\la-h_{\la'} + (A\omega^0,w_A(\gamma+\la-\la'+\rho))
+{g\over 2}|A\omega^0|^2 \eq $$
Since $\gamma+\la-\la'=\beta\in Q$, since $Q$ is invariant under $W$,
and since $(A\omega^0,\beta)\in\Zb$, the last two terms reduce
to (\omeg), i.e., to an integer.

The above calculation illustrates the dual relation between selection rules
and field identifications. If $\al\not =1$ the branching function vanishes
and, on the other hand, $\al=1$ implies that fields related by an
outer automorphism must be identified. However, the situation is not as
simple as in the integer level case, where $\al$ did not depend upon the
$\la$ fields. Then $\al$ was directly related to a difference between
elements of the covering group for the coset numerator and denominator.
Only the first part of the phase (\phaseA) appeared.
Nevertheless, we fall back to this situation once we know that the
fractional part is conserved in tensor product decompositions.

For unitary coset models, this is the whole story on field identifications.
However, in the nonunitary case, the problem is much richer, as we will
now see.

\subsubsection{Field identification in the fractional sector.}

The expression (\ScosetC) for the modular $S$ matrix depends only
on the integer parts of the weights, i.e. it does not explicitly
depend on $y$ or on $\la^{F,y}$.
This is an immediate source of field identification:
Two coset fields $\{\gamma,\la;\la'\}$ and $\{\xi,\mu;\mu'\}$ will be
identified if
$$ \left\{\eqalign{
&\gamma=\xi\com \la^I=\mu^I\com \la'^I=\mu'^I\cr
&\la^{F,y} = \mu^{F,y'}~\mod{Q^\vee} \cr}\right. \eqlabel\diffyId $$
Equality of the fractional parts modulo the coroot lattice $Q^\vee$
instead of the root lattice is required in order to preserve the
conformal dimension. This can be seen from (\dimCosI): a shift of $\la^{F,y}$
by an element of $Q^\vee$ (without modifying the other fields) does
not change $h$ modulo $\Zb$.
Such a shift does not affect the branching condition since
$Q^\vee\subset Q$.

A large class of field identifications can be obtained by assuming that
$\la^{F,y} = \mu^{F,y'}$ (strict equality) and $y\not=y'$.
Then $\mu=w.\la$ and $\mu' = w.\la'$, where $w=y'y^{-1}$.
Since $\la$ and $\la'$ have the same fractional part, they share the same
set $R^\la$ and the same associated subgroup $W^\la$. Thus if $w.\la$ is
admissible, so is $w.\la'$.
Furthermore, it is clear that if $\gamma$, $\la$ and $\la'$ satisfy the
branching condition, this condition is also satisfied by the weights
$\gamma$, $w.\la$ and $w.\la'$.
Finally, the conformal dimension (\dimmod) of the coset field is not affected
by a simultaneous shifted action of the Weyl group on any
weight of the coset field.
Thus, one has the identification
$$ \{\gamma,\la;\la'\} \sim \{\gamma,w.\la;w.\la'\} $$
This identification may be seen more easily from the following relation:
$$ S^{(m)}_{\mu\la}~S^{(m+l)}_{\mu'\la'}~=~
\phi^{(m+g)}_{\mu^I,\la}~\phi^{(m+g+l)*}_{\mu'^I,\la'}
e^{2\pi i(\mu^F,\la-\la')} \eq $$
which is manifestly invariant under
the simultaneous actions $\mu\to w.\mu$ and $\mu'\to w.\mu'$.

Field identifications obtained by the shifted action of the Weyl group
arise only in the fractional sector (i.e., the sector with
non-integer finite weights).
Indeed, we have shown that for a
given admissible weight $\la$, the number of elements $w$ such that
$w.\la$ is also admissible is $n(W)/n(W^\la)$ (including $w=1$).
For integrable weights $W^\la = W$, hence this yields no field identification.
Thus, coset fields occur in sets of
$n(W)/n(W^\la)$ fields identifiable with shifted actions of the Weyl group.

This class of field identification, together with
identification from outer automorphisms, appears to perform
all necessary field identifications for the classical algebras
$A_r$, $B_r$, $C_r$ and $D_r$.
This belief is based on the analysis of a large number
of examples, performed by computer.
For the algebras $A_r$ and $D_r$, we notice that all these identifications
occur within the $y=1$ sector, while this isn't true for $B_r$ and $C_r$.
On the other hand, for the exceptional algebra $G_2$,
it is only within the $y=1$ sector that all identifications are of the
above type. Otherwise, field identifications of the type (\diffyId) with
$\la^{F,y} \not= \mu^{F,y'}$ are necessary.
For all algebras studied, there is always a coset field representative
in the $y=1$ sector with $\lab^F=0$ (see next subsection).

A special case of identifications by shifted action of $W$ is
particularly useful for $\su{N}$ cosets.
Let us associate with each element $A\in O(\gh)$
an operator $B_A$ whose action on a weight $\la$ is defined by
$$ B_A\la := \l\{\eqalign{
&yw_Ay^{-1}.\la\qquad
\hbox{if the result is admissible with the same $y$}\cr
& \la \qquad {\rm otherwise}\cr }\r. \eqlabel\BAdef $$
Notice that
$$ yw_Ay^{-1}.\la = y.(A\la^I - (m+g)(A(\la^{F,y}+\omega^0)-\omega^0) \eq $$
This particular shifted action of $W$, together with outer automorphisms,
will be sufficient to perform all field identifications for $A_r$.

\subsubsection{A class of coset primary field representatives.}

The results of the last subsection suggest
that it  is possible to find a proper set of
coset primary fields characterized by a vanishing finite fractional part, i.e.,
fields of the form
$$\{\gamma,\la^I;\la'^I\}\qquad\hbox{with}\qquad
\bar\gamma+\lab^I-\lab'^I=0~\mod{Q}\com \eq $$
modulo the action of the outer automorphism group.
For $su(N)$, it turns out that it is always possible to choose inequivalent
$y$'s such that all $\la^{F,y}\in P^{u-1}_+$ (see appendix C).
If $y\not=1$, some
Dynkin labels must satisfy a stronger constraint than $\la^{F,y}_\mu\geq 0$.
As a result, for $su(N)$, the set of admissible $\la^{F,y}$ for $y\not=1$
is a proper subset of the set of admissible $\la^{F,1}$.
{}From (\diffyId),
it is therefore manifest in this case that all fields from the $y\not=1$
sectors can be identified with fields of the $y=1$ sector, and it is
sufficient to consider this sector only. Moreover, in the $y=1$ sector,
all fields with $\lab^{F,1}\not=0$ (the finite part) can be related to
fields with $\lab^{F,1}=0$ by using the operators $A$ and $B_A$.
This will be illustrated in the next section where canonical chains
of field identifications are constructed.

For other classical Lie algebras, the group of outer automorphisms is not
sufficiently large to relate all fields with $\lab^{F,y}\not=0$ to those
with $\lab^{F,y}=0$, even in the $y=1$ sector.
Furthermore, we cannot choose representative $y$'s in $W/W(A)$ such
that $\la^{F,y}\in P_+^{u-1}$, so that a priori we have no reason to
restrict ourselves to the $y=1$ sector.
Nevertheless, we can still use Eq.(\diffyId) to identify coset fields,
and for all examples we have considered it is possible to pick a set of
primary field representatives by restricting the search to $y=1$ and
$\lab^{F,1}=0$.

\section{Examples.}
\subsection{$\su2$ with $l=1$ and $m=-4/3$.}

This coset model has central charge $c=-\frac{22}5$ and, as shown by
Cardy [\Cardy], describes the Yang-Lee singularity.
It is the simplest nonunitary diagonal coset. Here $m^I=0$ so $\la^I$
vanishes.
The coset fields are then constructed from the triplets $\gamma_1=i$,
$\la_1=-(m+g)\la^F_1 = -\frac23 k$ and
$\la'_1=\la'^I_1-(m+g+1)\la^F_1 = n'-\frac53 k$ which satisfy the relation
$i+k-n' = 0\mod2$ (here $0\leq i\leq l=1$, $0\leq k\leq u-1=2$ and
$0\leq n'\leq m^I+u=3$).
There are 12 coset fields. They can be grouped in two sets according to
their conformal dimensions, given by (\dimmod).
Actually, $h_{\{\gamma,\la;\la'\}}\geq h_\gamma + h_\la - h_{\la'}$, and
generically the correct conformal dimension is the maximum of the latter
expression, taken over a set of identified coset fields.
Using the notation $\{\bar\gamma,\lab;\lab'\}$,
the coset fields of dimension 0, corresponding to
the identity, are
$$ \eqalign{
h=0~:~&\{0,0;0\}\com \{1,-\frac43;-\frac13\}\com\{1,-\frac23;-\frac53\}\com\cr
&\{0,-\frac23;\frac43\}\com\{0,-\frac43;-\frac{10}3\}\com\{1,0;3\}\cr }\eq $$
All other fields have dimension $h=-\frac15\mod1$:
$$ \eqalign{
 h=-\frac15~:~&\{1,0;1\}\com
\{0,-\frac43;-\frac43\}\com\{0,-\frac23;-\frac23\}\com\cr
&\{1,-\frac23;\frac13\}\com\{1,-\frac43;-\frac73\}\com \{0,0;2\}\cr}\eq $$
The equality of their conformal dimensions suggests that the fields in
each set can be identified (this is of course a necessary but not sufficient
condition for field identification).
The identification of the fields within each set can be established
by the action of the operators $a$ ($a[\la_0,\la_1]:=[\la_1,\la_0]$)
and $b:=B_a$ (see Eq.(\BAdef)).
It is very simple to check that, in each of the above two sets
and in the order shown, the fields are related by the following chain of
operators: $ababa$. For example,
$$ \{1,-\frac23;-\frac53\}~=~ba\{0,0;0\}\com
\{1,0;3\}~=~ababa\{0,0;0\} \eq $$
Notice that the above chain of identifications starts with a coset field
with $\lab^F=k=0$ and ends up with another field with $k=0$, on which
a further application of $b$ is neutral. Since the chain of operators
has 5 elements, one obtains $6~(=2u)$ field identifiactions.

\subsection{$\su2$ with $l=1$ and $m=-1/2$.}

In this example $u=2$ and $m^I=1$. One finds 16 fields which can be
grouped into four sets according to their conformal dimensions modulo 1:
$$\eqalign{
h=0~:~&\{0,0;0\}\com \{1,-\frac12;\frac12\}\com
\{1,-\frac32;-\frac52\}\com\{0,1;3\}\cr
h=-\frac1{20}~:~&\{1,0;1\}\com \{0,-\frac12;-\frac12\}\com
\{0,-\frac32;-\frac32\}\com\{1,1;2\}\cr
h=\frac15~:~&\{0,0;2\}\com \{1,-\frac12;-\frac32\}\com
\{1,-\frac32;-\frac12\}\com\{0,1;1\}\cr
h=\frac34~:~&\{1,0;3\}\com \{1,-\frac12;-\frac52\}\com
\{0,-\frac32;\frac12\}\com\{1,1;0\}\cr }\eq $$
One verifies that within each set all the fields can be identified
according to the chain of operators $aba$.
Again, starting with a field with $k=0$, one terminates on another
field with $k=0$.
Here we have $4~(=2u)$ field identifications.

\subsection{$\su3$ with $l=1$ and $m=-3/2$.}

This is the simplest $\su3$ diagonal nonunitary coset model, with
$u=2$ and $m^I=0$. For $su(3)$, the condition that $\mub=(\mu_1,\mu_2)\in Q$
is equivalent to
$$\eqalign{2\mu_1+\mu_2 ~=~0\mod3\cr \mu_1+2\mu_2 ~=~0\mod3\cr}\eq $$
By applying the above conditions one finds 18 coset fields in the $y=1$
sector (24 in all sectors)\foot{We have shown in appendix
C that, for $su(N)$, we
can always choose the $y$'s such that $\la^{F,y}\in P_+^{u-1}$, and, as
argued in section 4.3.4, this allows us to concentrate on the $y=1$
sector.}
which are grouped into two sets of
conformal dimensions 0 and $-\frac15$ (modulo 1).
In the notation $\{\bar\gamma,\lab;\lab'\}$, the states with $h=0\mod1$ are
$$\eqalign{
&\{(0,0),(0,0);(0,0)\}\com\{(1,0),(-\frac32,0);(-\frac12,0)\}\com
\{(0,1),(0,-\frac32);(0,-\frac12)\}\cr
&\{(0,1),(-\frac32,0);(-\frac52,0)\}\com
\{(0,0),(0,-\frac32);(2,-\frac52)\}\com\{(1,0),(0,0);(0,2)\}\cr
&\{(0,0),(-\frac32,0);(-\frac52,2)\}\com\{(1,0),(0,-\frac32);(0,-\frac52)\}
\com\{(0,1),(0,0);(2,0)\}\cr}\eq $$
They can be identified (in the above order) with the following chain:
$$ a^2ba^4ba^2 ~=~ [(a^2b)a^2]^2 \eq $$
(Recall that $a[\la_0,\la_1,\la_2]=[\la_2,\la_0,\la_1]$ and $b:=B_a$.)
This chain contains a loop since two fields in the middle of the chain
are reproduced twice ($a^3=1$).
This is due to the fact that the action of $b$ is neutral on
$a^2ba^2\La$, where $\La$ is a state with $\lab^{F,1}=0$.
However, it is not neutral on $aba^2\La$, and in order not to miss
the state $a^2ba^2\La$ a loop has to be introduced.
The chain contains 8 distinct elements, so that 9 fields can be
identified.
The same analysis holds for the fields with $h=-\frac15$.
We therefore found another coset model with two independent fields
(with $h=0$ and $h=-\frac15$). Its central charge is again $c=-\frac{22}5$,
and it yields another representation of the Yang-Lee singularity.

\subsection{Remark on the KNS duality for $\su{N}$ diagonal
cosets with $l=1$ and $m^I=0$.}

The fact that the Yang-Lee singularity can be desribed either by an $\su2$
or an $\su3$ diagonal coset has first been observed in [\Kuniba].
It is the simplest example of a general duality between $u$ and $N$
for $\su{N}$ diagonal cosets with $l=1$ and $m^I=0$. For such models
the central charge is
$$ c = -{(N-1)(u-1)(N+u+Nu)\over N+u} \eq $$
which makes manifest the $u\leftrightarrow N$ duality [\Kuniba,\ref{D.
Altschuler, M. Bauer and H. Saleur, \jr{J. Phys. A}\vol{23} (1990) 1789.}].
Hence $\su{N}$ diagonal cosets with $l=1$ and $m=N/u-N$ are equivalent to
$\su{u}$ diagonal cosets with $l=1$ and $m=u/N-u$. Note that the
condition $m^I=0$ fixes $t$ to be $t=-N(u-1)$. Since $u$ and $t$ must
be coprime, $u$ and $N$ must also be coprime.
The Yang-Lee singularity corresponds to the pair $(N,u)=(2,3)$.
The next simplest case is $(N,u)=(3,4)$, with $c=-\frac{114}7$.
(We have analyzed the corresponding two cosets and checked the equivalence
of the two models directly, from their modular $S$ matrices).
This duality has been proved under the assumption that it is possible
to consider only the $y=1$ sector with weights such that $\lab^{F,1}=0$.
That assumption is now justified by our analysis of field identifications.

\subsection{Remark on the number of field
identifications for $\su{N}$ cosets with $l=1$.}

As already pointed out, $\su{N}$ diagonal cosets with $l=1$ provide an
alternate description of minimal $W_N^{(p,p')}$ models, with $p=m^I+N$ and
$p'=p+u$.
By showing how both descriptions yield the same modular $S$ matrix, we have
established this result rigorously for all values of $p$ and $p'$.
In the Feigin-Fuchs representation, the primary fields are described by
two independent integrable weights $\la^I$ and $\la'^I$ of levels
$m^I$ and $m^I+u$ respectively, and the fields are identified according to
$$ (\la^I\mid\la'^I)~\sim~(a^i\la^I\mid a^i\la'^I)\com 1=1,\dots,N \eq $$
where $a$ denotes a cyclic permutation of the affine Dynkin labels.
Since the number of integrable fields at level $m^I$ is
$(m^I+N-1)!/(m^I!(N-1)!)$, the total number of inequivalent fields
(obtained by multiplying the above number by similar expression with
$m^I\rightarrow m^I+u$ and dividing by the order of the outer-automorphism
group) is
$$ {(m^I+N-1)!(m^I+N-1+u)!\over m^I!(m^I+u)!N!(N-1)!} \eqlabel\Na $$
In the coset description, there are $N$ possible values of $\gamma$, but
(\branchA) reduces the number of $\la'^I$ fields by a factor of $N$.
Thus, in the $y=1$ sector, there are
$$ {(m^I+N-1)!(m^I+N-1+u)!(m^F+N-1)!\over m^F!m^I!(m^I+u)![(N-1)!]^3}
\eqlabel\Nb $$
fields. Comparison of (\Na) and (\Nb) shows that there must be
$$ {(u+N-2)! N\over (u-1)!(N-1)!}\eqlabel\Niden $$
field identifications.

For $u=1$, this number reduces to $N$, as it should, being the order
of the automorphism group. For $\su2$, it reduces to $2u$ (compare
with sections 5.1 and 5.2) while for $\su3$ it is $\frac32 u(u+1)$
(compare with section 5.3).

The above argument is specific to $l=1$, since only then are there
$N$ different $\gamma$'s, and only then does the constraint (\branchA)
amount to restrict the number of fields by a factor of $N$.
Furthermore, there are no fixed points if $l=1$.

\subsection{Canonical chains of field identifications for
$\su{N}$ diagonal cosets.}

It is straightforward to generalize the chains of field identifications
obtained before for $\su2$ with $u=2,3$ and $l=1$.
For a general value of $u$ it is simply $(ab)^{u-1}a$, starting and
ending on states with $k=0$.
This yields $2u$ fields identified, in agreement with (\Niden).
Notice that apart from the extremities of the chain, there are no states
with $k=0$. This means that the coset fields with $k=0$ are always
identified according to
$$ \{\gamma,\la^I;\la'^I\} \sim (ab)^{u-1}a \{\gamma,\la^I;\la'^I\}\eq$$
Since on $\la^I$ and $\la'^I$ the action of $b$ is the same as that of $a$,
this can be rewritten as
$$ \{\gamma,\la^I;\la'^I\} \sim \{a^u\gamma,a\la^I;a\la'^I\}\eqlabel\idenA $$
Therefore, it is only for odd $u$ (and in particular for the unitary
case $u=1$) that the naive field identification
$ \{\gamma,\la^I;\la'^I\} \sim a \{\gamma,\la^I;\la'^I\}$ is sufficient.
This simple remark shows that it would not have been possible to
understand properly the field identifications by restricting ourselves
from the onset to fields with $\lab^F=0$.

It is also simple to work out the canonical chain of field identifications
for $\su3$ for arbitrary values of $u$. The result is
$$ \l[ \prod_{i=1}^{u-1}(a^2b)^i a^2\r] (a^2b)^{u-1}a^2\La \eq $$
where again we denote by $\La$ a field with $\lab^{F,1}=0$.
For $u=2$ this reduces to the chain found in section 5.3.
To count the number of distinct elements, one multiplies by 3 the
number of factors of $a^2b$, and adds 3 (2 for the first factor of
$a^2$ and 1 for the initial state).
The result is $\frac32 u(u+1)$, in agreement with (\Niden).
In each chain there are two $\La$ fields apart from the inital one,
namely $(a^2b)^{u-1}a^2\La$ and $[(a^2b)^{u-1}a^2]^2\La$.
For these fields the analog of (\idenA) is
$$ \{\gamma,\la^I;\la'^I\} \sim \{a^u\gamma,a\la^I;a\la'^I\}
\sim \{a^{2u}\gamma,a^2\la^I;a^2\la'^I\}\eq $$

As a final remark we display the canonical chain of field identification for
$\su{N}$ with $u=2$:
$$ [(a^rb)a^r]^r \La\com (r=N-1)\eq $$
This yields $N^2$ field identifications, in agreement with (\Niden).

It should be stressed that $\su{N}$ canonical chains depend only upon $u$,
and not on $m^I$ or $l$.
For special values of $l$ and $m^I$ there could be fixed points of $a^i$
(for some $i$), but this simply means that some chains will be truncated,
without affecting their completeness.
Such truncations can only arise from $a^i$ fixed points.
Indeed, as is easily seen,
states with $\lab^F\not=0$ cannot be neutral under the action of $b$
(this argument is specific to $su(N)$).
A necessary condition for an $su(N)$ coset field to be fixed under the action
of $b$ is that both $m^I$ and $u$ be multiples of $N$ (the dual Coxeter number
for $su(N)$). But this violates the requirement that $t$ and $u$ be coprime.

\subsection{$\su2$ with $l=2$ and $m=-4/3$.}

This is the simplest example of a non-unitary coset with $l\not=1$.
When $l$ is even the $\su2$ branching conditions are independent of $\la^F$.
All the coset fields can be grouped into six families as in table 2.

All the fields within each set can be identified, according to the chain
$(ab)^2a$, in the order above.
However, in the last set, it is truncated to $ab$, the state
$\{1,-\frac23;\frac13\}$ being a fixed point of $a$. This is possible because
$l$, $m^I$ and $u-1$ are all multiples of 2, the dual Coxeter of $su(2)$.

\vglue 12pt
{\baselineskip=16pt
\centerline{Table 2 : Coset fields of $\su2$ at $m=-4/3$ $(y=1)$.}
\nobreak
$$\vbox{\eightrm\halign{
\hfil#\hfil&\hfil#\hfil&\hfil#\hfil&\hfil#\hfil
&\hfil#\hfil&\hfil#\hfil\cr
\multispan6\hrulefill\cr
$h=0$ & $h=\frac12$ & $h=\frac14$ & $h=-\frac14$ &
$h=-\frac3{32}$ & $h=-\frac7{32}$ \cr
\multispan6\hrulefill\cr
$\{0,0;0\}$ & $\{2,0;0\}$ & $\{0,0;2\}$ & $\{2,0;2\}$ &
$\{1,0;1\}$ & $\{1,0;3\}$ \cr

$\{2,-\frac43;\frac23\}$ & $\{0,-\frac43;\frac23\}$ &
$\{2,-\frac43;-\frac43\}$ & $\{0,-\frac43;-\frac43\}$ &
$\{1,-\frac43;-\frac13\}$ & $\{1,-\frac43;-\frac73\}$ \cr

$\{2,-\frac23;-\frac83\}$ & $\{0,-\frac23;-\frac83\}$ &
$\{2,-\frac23;-\frac23\}$ & $\{0,-\frac23;-\frac23\}$ &
$\{1,-\frac23;-\frac53\}$ & $\{1,-\frac23;\frac13\}$ \cr

$\{0,-\frac23;\frac{10}3\}$ & $\{2,-\frac23;\frac{10}3\}$ &
$\{0,-\frac23;\frac43\}$ & $\{2,-\frac23;\frac43\}$ &
$\{1,-\frac23;\frac73\}$ & $$ \cr

$\{0,-\frac43;-\frac{16}3\}$ &
$\{2,-\frac43;-\frac{16}3\}$ & $\{0,-\frac43;-\frac{10}3\}$ &
$\{2,-\frac43;-\frac{10}3\}$ & $\{1,-\frac43;-\frac{13}3\}$ & $$ \cr

$\{2,0;6\}$ & $\{0,0;6\}$ & $\{2,0;4\}$ &
$\{0,0;4\}$ & $\{1,0;5\}$ & $$ \cr
\multispan6\hrulefill\cr}}$$
}\vglue 12pt

$\su2$ diagonal cosets with $l=2$ describe minimal superconformal
models, for which
$$ c = {3\over 2}\l(1 - 2{(p'-p)^2\over pp'}\r) \eqlabel\csuper$$
and
$$ h_{r,s} = {(rp'-sp)^2-(p'-p)^2\over 8pp'} + {1-(-)^{rs}\over 32} \eq $$
with $p'-p=2u$, $m+2=2p/(p'-p)$, $1\leq r\leq p-1$ and $1\leq s \leq p'-1$.
In the present case $p=2$, $p'=8$, and the superconformal fields have
dimensions
$$ h_{11}=0\com h_{13}=-\frac14\com h_{12}=-\frac3{32}
\com h_{14}=-\frac7{32}\eq $$
The last two lie in the Ramond sector ($r+s$ odd) while the first two lie
in the Neveu-Schwarz sector ($r+s$ even).
Fields in the NS sector are actually superfields whose component fields
have dimensions $h$ (given above) and $h+\frac12$.
In this way we recover the spectrum listed in table 2.

It is interesting to notice that the component fields in a given multiplet
are related to each other by the action of $a$ restricted to the field
$\gamma$. The two coset fields $\{\gamma,\la;\la'\}$ and
$\{a\gamma,\la;\la'\}$ form a NS multiplet if $a\gamma\not=\gamma$,
while fields in the Ramond sector are characterized by the condition
$a\gamma=\gamma$.

\subsection{$\so8$ with $l=1$ and $m=-5/2$.}

For this coset one finds 24 distinct fields, while the full $y=1$ sector
contains 384 fields. There are 16 field identifications, without any fixed
points. All these fields can be related by the action of $A$ and $B_A$.
With
$$\eqalign{
a[\la_0,\la_1,\la_2,\la_3,\la_4]~&=~[\la_1,\la_0,\la_2,\la_4,\la_3] \cr
\at[\la_0,\la_1,\la_2,\la_3,\la_4]~
&=~[\la_4,\la_3,\la_2,\la_1,\la_0] \cr} \eq $$
and $b:=B_a$, $\bt := B_{\at}$, the canonical chain of identification is
$$ a\at ab\at^2 a\at ba^2\at a\bt a\at a \eq $$
It starts and ends on an integral state (i.e., a state with $\lab^F=0$).
Furthermore, there are two intermediate integral states:
$ a\at a\bt a\at a \La$ and $\at a\at ba^2\at a\bt a\at a\La$, for a
total of four.
The integral states corresponding to the vacuum are
$$\eqalign{
& \{(0,0,0,0),(0,0,0,0);(0,0,0,0)\}\cr
& \{(0,0,0,0),(1,0,0,0);(3,0,0,0)\}\cr
& \{(0,0,0,0),(0,0,1,0);(0,0,3,0)\}\cr
& \{(0,0,0,0),(0,0,0,1);(0,0,0,3)\}\cr }\eq $$
They are related to each other by the chain $a\at a$ acting on
the integer parts $\la^I$ and $\la'^I$ only. In other words, coset fields
in the $\lab^F=0$ sector can be identified according to
$$ \{\gamma,\la^I; \la'^I\} ~\sim~\{A^u\gamma,A\la^I;A\la'^I\}
\eqlabel\idenLaF $$
\nobreak
\subsection{$\so5$ with $l=1$ and $m=-12/5$.}

This is again a coset with $m^I=0$. Since $B_2$ is not simply laced
there is a restriction on $u$: here $u\not\in 2\Zb$ (c.f. footnote $b$).
If, for simplicity's sake, we insist on having $m^I=0$, then $u$ and $g=3$
must be coprime, and the simplest example is then $u=5$.
One finds 18 distinct fields whose representatives can all be chosen
with $y=1$ and $\lab^F=0$. These are given in table 3.

The first 12 fields can be grouped into multiplets with $\Delta h=\hf$.
In fact, the two fields in a multiplet are related by the action of $a$
on $\gamma$ (for $B_r$, the only non-trivial outer automorphism
is $a[\la_0,\la_1,\dots,\la_r] = [\la_1,\la_0,\dots,\la_r]$).
Similarly, the remaining 6 fields are characterized by the
property $a\gamma=\gamma$. One thus finds NS and R sectors, exactly as in
the superconformal case [\Fateev].

\vglue12pt
{\baselineskip=16pt
\centerline{Table 3 : Distinct coset fields of $\so5$ at $m=-12/5$.}
\nobreak
$$\vbox{\eightrm\halign{
\hfil#\hfil&~\hfil#\hfil&\qquad\hfil#\hfil&~\hfil#\hfil\cr
\multispan4\hrulefill\cr
$\qquad h\qquad$ & $\{\bar\gamma,\lab^I;\lab'^I\}$ & $\qquad h\qquad$ &
$\{\bar\gamma,\lab^I;\lab'^I\}$\cr
\multispan4\hrulefill\cr
0 & $\quad\{(0,0),(0,0);(0,0)\}\quad$ &
$-\frac14$ & $\quad\{(1,0),(0,0);(1,2)\}\quad$\cr
$\frac12$ & $\{(1,0),(0,0);(0,0)\}$ &
$-1$ & $\{(1,0),(0,0);(0,4)\}$\cr
$-\frac54$ & $\{(0,0),(0,0);(1,0)\}$ &
$-\frac12$ & $\{(1,0),(0,0);(0,4)\}$\cr
$-\frac34$ & $\{(1,0),(0,0);(1,0)\}$ &
$-\frac{15}{32}$ & $\{(0,1),(0,0);(0,1)\}$\cr
$-\frac78$ & $\{(0,0),(0,0);(0,2)\}$ &
$-\frac{23}{32}$ & $\{(0,1),(0,0);(0,5)\}$\cr
$-\frac38$ & $\{(1,0),(0,0);(0,2)\}$ &
$-\frac{31}{32}$ & $\{(0,1),(0,0);(0,3)\}$\cr
$-\frac98$ & $\{(0,0),(0,0);(2,0)\}$ &
$-\frac{33}{32}$ & $\{(0,1),(0,0);(1,1)\}$\cr
$-\frac58$ & $\{(1,0),(0,0);(2,0)\}$ &
$-\frac{37}{32}$ & $\{(0,1),(0,0);(1,3)\}$\cr
$-\frac34$ & $\{(0,0),(0,0);(1,2)\}$ &
$-\frac{39}{32}$ & $\{(0,1),(0,0);(2,1)\}$\cr
\multispan4\hrulefill\cr}}$$
}\vglue 12pt
\vglue12pt
\centerline{Table 4 : coset fields identified to $\{(1,0),(0,0);(1,0)\}$.}
\nobreak
$$\vbox{\eightrm\halign{
#\hfil&\qquad#\hfil\cr
\multispan2\hrulefill\cr
$\{(1,0),(0,0);(1,0)\}$&
$\{(0,0),(-\frac{12}5,0);(-\frac{12}5,0)\}$\cr
$\{(0,0),(-\frac35,0);(-\frac35,0)\}$&
$\{(1,0),(-\frac95,0);(-\frac45,0)\}$\cr
$\{(1,0),(-\frac65,0);(-\frac{11}5,0)\}$&
$\{(0,0),(-\frac65,0);(\frac45,0)\}$\cr
$\{(0,0),(-\frac95,0);(-\frac{19}5,0)\}$&
$\{(1,0),(-\frac35,0);(\frac{12}5,0)\}$\cr
$\{(1,0),(-\frac{12}5,0);(-\frac{27}5,0)\}$&
$\{(0,0),(0,0);(4,0)\}$\cr
\multispan2\hrulefill\cr
$\{(1,0),(0,-\frac65);(1,-\frac{16}5)\}$&
$\{(0,0),(-\frac65,-\frac65);(\frac45,-\frac{16}5)\}$\cr
$\{(0,0),(-\frac35,-\frac65);(-\frac35,-\frac{16}5)\}$&
$\{(1,0),(-\frac35,-\frac65);(\frac{12}5,-\frac{16}5)\}$\cr
$\{(1,0),(-\frac65,-\frac65);(-\frac{11}5,-\frac{16}5)\}$&
$\{(0,0),(0,-\frac65);(4,-\frac{16}5)\}$\cr
$\{(0,0),(-\frac{12}5,0);(4,-\frac{32}5)\}$&
$\{(1,0),(0,-\frac{12}5);(1,-\frac{32}5)\}$\cr
\multispan2\hrulefill\cr
$\{(0,0),(0,-\frac{12}5);(4,-\frac{32}5)\}$&
$\{(1,0),(0,-\frac{12}5);(1,-\frac{32}5)\}$\cr
\multispan2\hrulefill\cr}}$$
\vglue 12pt

Let us now turn to the question of field identification.
In the $y=1$ sector, the field $\{(1,0),(0,0);(1,0)\}$ is identified to
17 other fields. (This is true of all NS fields; in the R sector, there are
fixed points.)
The 18 identified fields can be divided into three groups, as shown in
table 4.

Within each group, the fields are identified using $a$ and $b=B_a$.
The chains of field identification are respectively $(ab)^4a$, $(ab)^2a$
and $a$.
Now, the different groups can be identified according to (\diffyId).
Indeed, the last coset field of the second group, and the first of the third
group have the same $\bar\gamma$, $\lab^I$ and $\lab'^I$, and their
$\lab^{F,1}$ differ by $(0,2)=2\al_2+\al_1\in Q^\vee$.
Therefore they are not distinct coset fields.
The same thing applies to the last field of the first group and the first
of the last group, with $\Delta\lab^{F,1} = (0,4)\in Q^\vee$.
Here again we observe that in the $\lab^F=0$ sector, fields are identified
according to (\idenLaF)

\subsection{A remark on ${\hat B}_r$ diagonal cosets.}

For diagonal ${\hat B}_r$ (or $\so{2r+1}$) cosets with $l=1$, the splitting
into NS and R sectors is a general feature. The central charge is
$$ c = (r+\hf)\l\{ 1 - {2r(2r-1)\over (m+2r)(m+2r-1)}\r\}\com \eq $$
and we recover the superconformal minimal series for $r=1$
(set $m+1=p/(p'-p)$).
This however is purely formal since ${\hat B}_1$ does not really make
sense.
On the other hand, for $r\geq 2$, although one still has the analog of NS and
R sectors, these sectors are characterized
by the value of $\gamma$. When $l=1$, $\gamma$ can take only three values,
namely $\omega^0$, $\omega^1$ and $\omega^r$. The first two are related
by the action of $a$ and correspond to the NS sector.
Since the $r$th Dynkin label of any root is even, the condition
(\branchA) implies $\la_r+\la'_r\in 2\Zb$ in the NS sector
(where $\gamma_r=0$), while $\la_r+\la'_r\in2\Zb+1$ in the R sector
(where $\gamma_r=1$).

\section{Conclusion.}

Highest weights for admissible representations of \KM algebras at
fractional level $m=t/u$ are of the form
$y\bullet(\la^I-(m+g)\la^{F,y})$ with $\la^I\in P^{m^I}_+$,
$m^I=u(m+g)-g$ and $\la_\mu^{F,y}\in r_\mu\Zb$.
For diagonal cosets with one factor in the numerator being at integer
level, field identifications allow us to restrict the choice of coset
field representatives to a fundamental sector: $y=1$ and $\lab^{F,y}=0$.
This fact, demonstrated for $\su{N}$, has been verified for other
algebras with a large number of
examples.\foot{The calculations have been carried out with the help
of a computer program performing field identifications with the
modular $S$ matrix. Examples from $B_{2-4}$, $C_{3-4}$, $D_4$ and $G_2$
have been analyzed.}
The independence of the coset modular matrix $S$ on $y$ and $\lab^F$
ensures that fields in the fundamental sector transform among themselves.
In the fundamental sector, fields can be identified according to
$$ \{\gamma,\la^I;\la'^I\} ~\sim~
\{A^u\gamma,A\la^I;A\la'^I\}\eqlabel\idy$$
where $\gamma\in P_+^l$, $\la^I\in P_+^{m^I}$, $\la'^I\in P_+^{m^I+lu}$ and
where $A$ is any element of the outer automorphism group. The branching
condition is $\bar\gamma + \lab^I-\lab'^I\in Q$.
This has been pointed out in all the examples displayed.
The following proof makes it a general result:
It is simple to see that
$$ AS_{\lab^I,\mub^I}^{(m)} =
e^{-2\pi i(A\omega^0, u\mu^I+(u-1)g)}
S_{\lab^I,\mub^I}^{(m)} \eq $$
Therefore, acting respectively with $A$, $A'$ and $A''$ on the product
$S_{\gamma,\xi}^{(l)}~S_{\lab^I,\mub^I}^{(m)}~S_{\lab'^I,\mub'^I}^{(m+l)}$
produces the phase
$$ \exp-2\pi i[(A\omega^0,\xi)+(A'\omega^0,u\mu^I+(u-1)\rho)
-(A''\omega^0,u\mu'^I+(u-1)\rho)] \eq $$
Canceling the factors containing $\rho$ requires $A'=A''$.
Then, using $\bar\xi + \lab^I-\lab'^I\in Q$, one needs $A = (A')^u$
in order to get a vanishing phase.

That nonunitary diagonal cosets can be described solely in terms
of weights whose finite parts are integrable, with fields identified according
to (\idenLaF), is the main result of this paper.
\vskip 12pt
\par\leftline{\tenbf Acknowledgements}\par\nobreak
P.M. would like to thank RIMS for its hospitality in the course of the
workshop {\it Infinite analysis}, where part of this work was done.
M.W. similarly thanks the CERN theory division.
We acknowledge useful discussions with C. Ahn and T. Nakanishi.
This work is supported by NSERC (Canada) and F.C.A.R. (Qu\'ebec).
\vskip 12pt
\par\leftline{\tenbf References}\par\nobreak
\immediate\closeout\refs \vskip 0.5cm
  \message{References}\input references
\tenrm

\appendix{Appendix A}

In this appendix we briefly describe the simple roots, coroots, the
Weyl groups and the outer automorphisms associated with the four classical
algebras $A_r$, $B_r$, $C_r$ and $D_r$.
In each case it is useful to introduce an orthogonal basis $\{e_i\}$
for weight space. The description of the roots and of the action of
the Weyl group $W$ is thus simplified. The conventional ordering defined
on the roots becomes a simple lexicographic ordering in this basis.
The group of outer automorphisms is isomorphic to the symmetry group
of the affine Dynkin diagrams.

\subsection{$A_r$ or $su(r+1)$.}

The orthogonal basis spans a space of dimension $r+1$, and the weight
space of $A_r$ is orthogonal to the vector $e_1+e_2+\dots+e_{r+1}$.
The positive roots are $e_i-e_j$, with $i<j$. The simple roots are
$\alb_i = e_i-e_{i+1}$, with $i=1,2,\dots r$.
The highest root is
$$\thb = e_1 - e_{r+1} = \sum_{i=1}^r \alb_i \eq $$
Therefore all the marks $k_i$ are equal to 1.
The algebra is simply laced, and the coroots coincide with the roots,
the comarks $k^\vee_i$ with the marks.
On a weight $\la=\sum_i^{r+1} \la^i e_i$, the effect of a Weyl reflection
with respect to $\alb = e_i-e_j$ is to transpose the components $\la^i$ and
$\la^j$. The Weyl group generated by these reflections is the permutation
group of the $r+1$ components $\la^i$.

The outer automorphisms are cyclic permutations of the simple roots,
generated by $a(\al_\mu) = \al_{\mu+1}$, $a(\al_r) = \al_0$. From the
expression for $\alb_i$ and $\thb$, we see that the corresponding Weyl
group element is $w_a$ such that $w_a(e_i)=e_{i+1}$, $w_a(e_{r+1}) = e_1$.
The subgroup $W(A)$ is isomorphic to $\Zb_{r+1}$.

\subsection{$B_r$ or $so(2r+1)$.}

The orthogonal basis spans a space of dimension $r$.
The positive roots are $e_i\pm e_j$, with $i<j$, and $e_j$.
The simple roots are $\alb_i = e_i-e_{i+1}$, with $i=1,2,\dots r-1$,
and $\alb_r = e_r$.
The highest root is
$$\thb = e_1 + e_2 = \alb_1 +2\alb_2 + 2\alb_3+\dots + 2\alb_r \eq $$
Therefore all the marks $k_i$ are equal to 2, except $k_1=1$.
The algebra is not simply laced, and $\alb^\vee_i = \alb_i$ except for
$\alb^\vee_r = 2\alb_r$. Similarly, $k^\vee_i=k_i$ except for
$k^\vee_r = \hf k_r = 1$.
As before, the effect of a Weyl reflection
with respect to $\alb = e_i-e_j$ is to transpose the components $\la^i$ and
$\la^j$.
But a Weyl reflection with respect to $\alb = e_i+e_j$ transposes the
components $\la^i$ and $\la^j$ and changes their sign, while
a Weyl reflection with respect to $\alb = e_i$ only changes the sign of
$\la^i$.
The Weyl group generated by these reflections is therefore the group of
$r!$ permutations of the $r$ components $\la^i$, associated with the $2^r$
possible sign changes of the components, for a total of $2^r r!$ elements.

The only non-trivial outer automorphism is a transposition of the
first two affine simple roots:
$a(\al_\mu) = \al_{\mu}$ except for $a(\al_0) = \al_1$ and $a(\al_1) = \al_0$.
The corresponding Weyl group element is $w_a$ such that $w_a(e_i)=e_i$
except for $w_a(e_1) = -e_1$.
The subgroup $W(A)$ is isomorphic to $\Zb_2$.

\subsection{$C_r$ or $sp(r)$.}

The Cartan matrix of $C_r$ is just the transpose of that of $B_r$, i.e.
the long roots and the short roots are interchanged.
The positive roots are $(e_i\pm e_j)/\sqrt{2}$, with $i<j$, and $\sqrt{2}e_j$.
The simple roots are $\alb_i = (e_i-e_{i+1})/\sqrt{2}$, with $i=1,2,\dots r-1$,
and $\alb_r = \sqrt{2}e_r$.
The highest root is
$$\thb = \sqrt{2}e_1 = 2\alb_1 +\dots + 2\alb_{r-1} + \alb_r \eq $$
Therefore all the marks $k_i$ are equal to 2, except $k_r=1$.
The algebra is not simply laced, and $\alb^\vee_i = 2\alb_i$ except for
$\alb^\vee_r = \alb_r$. Similarly, $k^\vee_i=1$, for all $i$.
The Weyl group is exactly the same as for $B_r$.

The only non-trivial outer automorphism amounts to reversing the order
of the affine simple roots:
$a(\al_\mu) = \al_{r-\mu}$.
{}From the expression for $\alb_i$ and $\thb$, we see that
the corresponding Weyl group element is $w_a$ such that $w_a(e_i)=-e_{r+1-i}$.
The subgroup $W(A)$ is again isomorphic to $\Zb_2$.

\subsection{$D_r$ or $so(2r)$.}

The orthogonal basis spans a space of dimension $r$.
The positive roots are $e_i\pm e_j$, with $i<j$.
The simple roots are $\alb_i = e_i-e_{i+1}$, with $i=1,2,\dots r-1$,
and $\alb_r = e_{r-1}+e_r$.
The highest root is
$$\thb = e_1 + e_2 = \alb_1+2\alb_2+\dots+2\alb_{r-2}+\alb_{r-1}+\alb_r \eq$$
Therefore all the marks $k_i$ are equal to 2, except $k_1=k_{r-1}=k_r=1$.
The algebra is simply laced, and $\alb^\vee_i = \alb_i$, $k^\vee_i=k_i$
for all $i$.
The Weyl group is similar to that of $B_r$, except that the sign changes must
always occur in pairs, since the vectors $e_i$ are not roots. The order of
the Weyl group is therefore $2^{r-1} r!$.

The group of outer automorphisms is not the same depending if $r$ is even or
odd. A generator common to the two cases is $a$, which
transposes $\al_0$ with $\al_1$, and  $\al_r$ with $\al_{r-1}$.
The corresponding Weyl group element is $w_a(e_i)=e_i$ except for
$w_a(e_1) = -e_1$ and $w_a(e_r) = -e_r$.
The number of sign changes is thus even.
If $r$ is even, we define $\at$ as $\at(\al_\mu)=\al_{r-\mu}$, and the
corresponding Weyl group element is $w_{\at}(e_i) = -e_{r+1-i}$.
Note that the number of sign changes is even, and that $\at^2=a^2=1$.
The subgroup $W(A)$ is thus isomorphic to $\Zb_2\times\Zb_2$.

If $r$ is odd, we define $\at$ as $\at(\al_\mu)=\al_{r-\mu}$, except for
$\at(\al_0)=\al_{r-1}$, $\at(\al_1)=\al_r$, $\at(\al_{r-1})=\al_1$ and
$\at(\al_r)=\al_0$.
The corresponding Weyl group element is $w_{\at}$ such that
$w_{\at}(e_i) = -e_{r+1-i}$, except for $w_{\at}(e_1) = e_r$. In this case
the number of sign changes is still even, with $\at^4=1$ and $\at^2 = a$.
The subgroup $W(A)$ is thus isomorphic to $\Zb_4$.

\appendix{Appendix B}

In this appendix we demonstrate that admissible weights satisfying the
conditions (\YLA-\LAFY) automatically satisfy the defining condition (\admiA).

Firstly, let us show that (\admiA $~ii$) is satisfied. The coroot
$\alv=(\alvb_\mu,n,0)$ ($n\in r_\mu\Zb$) will belong to $R^\la$ if
$$ (\alv,\la)\in\Zb \qquad{\rm or}\qquad
nt+j\in u\Zb\qquad(m=t/u) \eqlabel\NTJ $$
where $(\la,\alvb_\mu) \equiv j/u$.
According to (\YLA-\LAFY), $j$ is an integer in all cases. Moreover,
$j\in r_\mu\Zb$ if $(r_\mu,u)\not=1$, as we can check from (\YLA).
$j$ being an integer, it is always possible to find an integer solution $n$ to
(\NTJ), and this because $(t,u)=1$.\foot{
Recall that if $(t,u)=1$, then $\exists a,b\in \Zb$ such that $at+bu=1$.}
The set of solutions to (\NTJ) is then $n+u\Zb$. If $(r_\mu,u)=1$, there
will always be a member $n'$ of that set such that $n'\in r_\mu\Zb$, and
then the coroot $(\alvb_\mu,n',0)\in R^\la$.
If, on the other hand, $(r_\mu,u)\not=1$, i.e. if $u = ar_\mu$ with
$a\in\Zb$, then $j=j' r_\mu$ and (\NTJ) can be rewritten as
$$ (n/r_\mu)t+j'\in a\Zb \eq $$
and there is always an integer solution $(n/r_\mu)$ to this equation.
Therefore, if $\la$ satisfies the conditions (\YLA-\LAFY), the set $R^\la$
contains coroots of the form $(\alvb_\mu,n,0)$ ($n\in r_\mu\Zb$) for all
$\mu$, and consequently the rational span of $R^\la$ coincides with the
rational span of the simple roots.
Note that this proof is considerably simplified if all roots are long
roots, i.e. if $r_\mu=1~\forall\mu$.

Secondly, let us show that (\admiA $~i$) is satisfied.
Consider all the positive coroots given by
$\al_\mu^\la \equiv \la^{F,y}_\mu\ct + y(\alv_\mu)$. This set of $r+1$
roots has the same Cartan matrix as the $r+1$ original simple coroots.
However, the sum
$\sum_\mu k^\vee_\mu\al_\mu^\la$ is equal to $u\ct$,
i.e. the corresponding canonical central element is a multiple of $\ct$.
Therefore, any positive coroot $\alv$ is
of the form $\sum_\mu l_\mu\al_\mu^\la - k\ct$, where $0\leq k<u$ and
$l_\mu\in \Zb_+$. If we substitute this form into the condition (\admiA $~i$),
we obtain
$$(\la+\rho,\alv) = \sum_\mu(1+\la^I_\mu)l_\mu - (m+g)k\eq$$
If $k=0$, this is certainly not an element of $-\Zb_+$ for any root
$\al$, since $\la^I_\mu\geq 0,~\forall\mu$.
If $0<k<u$, then this is not an integer.
Therefore, any weight satisfying (\YLA-\LAFY) also satisfies the
conditions  (\admiA).

\appendix{Appendix C}

In this appendix we show that, in the case of $su(N)$, it is possible
to choose a representative $y$ in each class of $W(\gb)/W(A)$ such that
$\la^{F,y}\in P_+^{m^F}$.
Let us rewrite the condition (\LAFY):
$$ \ct \la^{F,y}_\mu + y\big(\al^\vee_\mu\big) \in R_+\eqlabel\CLAFY$$
Since positive coroots have a non-negative grade, it is clear that
the only negative value of $\la^{F,y}_\mu$ allowed is $-1$, and this only if
$\mu=0$.
In that case the l.h.s. of (\CLAFY) is simply $-y(\thb)$.
If $y(\thb)\in R_+$, then the value $\la^{F,y}_0 = -1$ is forbidden and
$\la^{F,y}\in P_+^{m^F}$.
If $y(\thb)\not\in R_+$, the question is if we can find an element
$w_a\in W(A)$ such that $yw_a(\thb)\in R_+$.
In the case of $su(N)$, $\thb = e_1-e_n$ (orthogonal basis, $n=r+1$),
and $y(\thb) = e_{y(1)}-e_{y(n)}$, where the action of $y$ on indices is
defined by the corresponding permutation.
If $y(1)>y(n)$, then $y(\thb)\not\in R_+$.
However, there is a suitable cyclic permutation $w_a$ such that
$yw(1)<yw(n)$, i.e., such that $yw(\thb)\in R_+$.
Therefore, by right application of a suitable $w_a\in W(A)$, we can make
sure that $\la^{F,y}\in P_+^{m^F}$.
It is easy to verify that this argument works uniquely in the case of
$su(N)$. The outer automorphism groups of other classical algebras are not
large enough to allow for such representative $y$'s.

\appendix{Appendix D}

In this appendix we show that if $\la$ is admissible, then for
any $w\in W$, there is a unique $\wt\in W^\la$ such that $w\wt.\la$
is also admissible.
Let $R^\la_+\subset R_+$ be the set of positive roots also part of
$R^\la$ (see (\Rlambda)).
Consider an element $w\in W$ such that $w(R^\la_+)\subset R_+$.
Since $\la$ is admissible (\admiA) is satisfied, and consequently
$$ (w.\la+\rho,w(\alb^\vee)) + mn \not\in -\Zb_+ \eq $$
where $\al^\vee = (\alb^\vee,0,n)$ is a positive coroot ($n>0$, or
$n=0$ with $\alb^\vee>0$).
If  $\al^\vee\in R^\la_+$, then by hypothesis $w(\al^\vee)\in R_+$
and $w.\la$ passes the test of admissibility for these coroots.
To prove that $w.\la$ satisfies (\admiA), it remains to consider
the case where $n=0$, $\al^\vee\not\in R^\la$ and $w(\alb^\vee)\in R_-$.
But in this case $-w(\alb^\vee)\in R_+$ and
$$-(w.\la+\rho,w(\alb^\vee)) = -(\la+\rho,\alb^\vee) \not\in\Zb\eq$$
and therefore (\admiA) is also satisfied.
Thus we have shown that $w.\la$ is admissible.
(The second condition of (\admiA) is trivially satisfied).

Next, we will show that for any $w\in W$ there is a unique
element $\wt\in W^\la$ such that $w\wt(R^\la_+)\subset R_+$.
Let us denote by $\pi$ the projection that takes $R$ into $R^\la$.
Then consider the set $\pi w^{-1}(R_+)\subset R^\la$.
This set is the image of $R^\la_+$ by some unique element $\wt\in W^\la$:
$$ \pi w^{-1}R_+ = \wt\pi R_+ \eq $$
Then $\wt(R^\la_+)\subset w^{-1}(R_+)$ and our assertion follows.

Finally, let us show that if $\la$ is an admissible weight
admitting a decomposition (\YLA) with an element $y$ of $W$, and
if $w.\la$ is also admissible, then $w.\la$ admits a decomposition
(\YLA) with $y$ replaced by $wy$, and $\la^{F,wy}=\la^{F,y}$.
This almost seems obvious, but we must check that (\LAFY) implies
$$ \ct\la^{F,y}_\mu + wy(\al^\vee_\mu) \in R_+ \eq $$
A problem might arise only when $\la^{F,y}_i = (\la^F,y(\al_i)) = 0$,
or when $\la^{F,y}_0 = (\la^F,y(\al_0)) = -1$.
In both cases one easily checks that
$y(\al_\mu)\in R^\la_+$, and so is $wy(\al_i)$ (by hypothesis).
Hence (\LAFY) is satisfied with $y\rightarrow wy$.

\bye